\documentclass[prd,aps,nofootinbib,onecolumn,floatfix,10pt]{revtex4}
\usepackage{amsmath,graphicx,color,epsfig}

\begin{document}
\title{Production of Charmed Tetraquarks from $B_c$ and $B$ decays}

\author{Xiao-Gang He$^{1,2,3}$~\footnote{Email:hexg@sjtu.edu.cn}, Wei Wang$^{1,4}$~\footnote{Email:wei.wang@sjtu.edu.cn} and Rui-Lin Zhu$^{1,4}$~\footnote{Email:rlzhu@sjtu.edu.cn}}
\affiliation{
$^1$ INPAC, Shanghai Key Laboratory for Particle Physics and Cosmology, Department of Physics and Astronomy, Shanghai Jiao-Tong University, Shanghai, 200240,   China\\
$^2$ CTS, CASTS and Department of Physics, National Taiwan University, Taipei, 10617, Taiwan, \\
$^3$ National Center for Theoretical Sciences, Hsinchu, 300, Taiwan\\
$^4$ State Key Laboratory of Theoretical Physics, Institute of Theoretical Physics, Chinese Academy of Sciences, Beijing 100190, China
 }

\begin{abstract}
Hadronic states composed of multi-quark flavors  may  exist in reality  since  they are  not prohibited by QCD. Compact four quark systems of color singlet are classified as  tetraquarks.  To understand the properties of these states, more theoretical and experimental efforts are needed. In this work, we study charmed tetraquarks with three light flavors using flavor $SU(3)$ symmetry. States with three different light quarks must be in a ${\bf \bar 6 }$ or a ${\bf 15}$ multiplet. We  investigate the production  of  charmed tetraquarks $X_c$ in $B\to X_c (\overline {X}_c) P$ and $B_c \to X_c P$ decays. Whether the states with three light quarks belong to ${\bf \bar 6}$ or ${\bf 15}$ can be determined by studying various  tetraquark $B$ and $B_c$ decays.  We demonstrate that the  decay amplitudes for these decays can be parametrized by a few irreducible SU(3) invariant amplitudes. We then derive relations for decay widths and CP violating rate difference which can be examined  experimentally.  Although no experimental measurement is available yet, they might be accessed at the ongoing and forthcoming experiments like the LHCb and Belle-II.  Measurements of these observables  can not only provide useful information for the study  of  exotics spectroscopy but are also valuable information towards a better understanding of  some non-perturbative aspects of QCD.
\end{abstract}
\maketitle

\section{Introduction}

Although most hadrons observed in experiments can be well accommodated in  quark model where mesons are composed of a quark and an anti-quark and baryons are made of three quarks, it is widely believed that  there exist structures beyond   naive quark model. They are generally called  as hadron exotics. In the past decades,  great progresses have been made  in  finding exotic hadron states.
A milestone  is   the discovery of $X(3872)$, firstly in $B$ decays by Belle~\cite{Choi:2003ue} and subsequently confirmed in many reactions by different experiments~\cite{Acosta:2003zx,Aubert:2004ns,Abazov:2004kp}.  Since then  the identification of the exotic hadrons becomes  a key topic in hadron physics.  A number of new interesting structures were discovered in the mass region of heavy quarkonium, now generically  under the name  XYZ states. For  recent reviews,   see Refs.~\cite{Brambilla:2010cs,Esposito:2014rxa,Chen:2016qju,Ali:2016gli}.

Many of the newly discovered XYZ states  defy the quark model assignment, meson as $\bar qq$  or baryon as $qqq$. Two natural  interpretations in theory include  the hadron molecule   in which XYZ states are formed by two  loosely bounded constituents,  and the compact tetraquark.
However  the determination of their internal degrees of freedom may be vague because   their quantum numbers can overlap with ordinary mesons and baryons. Thus looking for the four-quark states with different flavors are of unambiguously importance.
The D0 collaboration  has announced  the discovery of a bottomed state $X(5568)$ decaying into the $B_s\pi^\pm$~\cite{D0:2016mwd}.
Unfortunately, soon after the D0 announcement, the LHCb collaboration reported negative results in their search~\cite{LHCb:2016ppf}.
The existence of the $X(5568)$  claimed by  D0 was not supported  by the LHCb data.
Nevertheless, the possibility of such exotic state has  attracted a lot of theoretical attentions~\cite{Agaev:2016mjb,Wang:2016tsi,Zanetti:2016wjn,
Wang:2016mee,Chen:2016mqt,Agaev:2016ijz,Xiao:2016mho,Liu:2016xly,Liu:2016ogz,
Agaev:2016lkl,Dias:2016dme,Wang:2016wkj,Agaev:2016urs,He:2016yhd,Jin:2016cpv,
Stancu:2016sfd,Burns:2016gvy,Tang:2016pcf,Guo:2016nhb,Lu:2016zhe,Esposito:2016itg,
Albaladejo:2016eps,Albaladejo:2016hae,Ali:2016gdg,Wang:2016cel,Albuquerque:2016nlw,Chen:2016npt,Cheng:2015vpn,Wu:2016vtq,Terasaki:2016zbt,Chen:2016jxd,Agaev:2016srl}. The properties of such exotic states are still  being actively studied.

In this work, we propose to search for tetraquarks composed of a charm quark and three different light quarks states,  $\bar c d s \bar u$, $\bar c s u \bar d$, and $\bar c ud \bar s$, in  $B_c$ and $B$ decays. Whether such states exist will be subject to the future experimental investigations.  We first use  flavor SU(3) symmetry to classify the charmed tetraquarks $X_c$, and then estimate their masses by making use of a constituent quark model.  Finally we  explore  $B_c$ and $B$  decay into a $X_c$ state and a pseudoscalar octet state $P$  from flavor $SU(3)$ symmetry respectively. In particular we derive relations for  decay widths and CP violations among different decay channels, which can be tested by experiments.

This paper is organized as follows. In Sec.~\ref{sec:spectroscopy}, we give an overview  of $SU(3)$ classification of charmed tetraquarks with different light quarks and their associated members, and estimate their masses. In Sec.~\ref{sec:effective_hamiltonian} we study the $B_c(B) \to X_c P$ decay amplitudes using $SU(3)$ symmetry.  In Sec.~\ref{sec:discussions} we discuss some useful relations  for decay widths and CP violations in $B_c(B) \to X_c P$ decays. In the last section, we provide a brief summary of this work.

\section{Charmed tetraquarks spectroscopy  }
\label{sec:spectroscopy}

\subsection{Charmed tetraquarks $X_c$ in SU(3)}

Charmed tetraquark states $X_c \sim [qq']\bar{q}''\bar{c}$ with three light quarks, $q$, $q'$ and $q''$, can be conveniently organized by flavor $SU(3)$ symmetry~\cite{He:2016yhd}.
Under the flavor $SU(3)$ symmetry, the three light quarks, $(u,\; d,\; s)$ form a triplet ${\bf 3}$ representation and the charm quark $c$ is a singlet~\cite{Zeppenfeld:1980ex,Chau:1990ay,Gronau:1994rj}. Tetraquark states formed by three light quarks ($q$, $q'$ and $q''$ are different ones) and one charm quark can have the following irreducible representations
\begin{eqnarray}
{\bf 3}\otimes {\bf 3} \otimes  {\bf\bar 3} = {\bf3}\oplus {\bf3} \oplus {\bf\bar 6}\oplus {\bf15}\;.
\end{eqnarray}

We will be interested in charmed tetraquark states with four different quarks, namely $\bar c d s \bar u$, $\bar c s u \bar d$, and $\bar c ud \bar s$ states. They must be in a ${\bf \bar 6}$ or a ${\bf 15}$ representation.  We label ${\bf \bar 6}$ representation by $X_{[i,j]}^k$. Here the flavor components are antisymmetric under the exchange of i and j, and traceless $X^i_{[i,j]} = 0$. More explicitly,  the components are given by~\cite{He:2016yhd}
\begin{eqnarray}
 &&X_{[2,3]}^1= \frac{1}{\sqrt 2} X_{ds\bar u}', \;\;\;  X_{[3,1]}^2 = \frac{1}{\sqrt 2} X_{su\bar d}', \;\;\; X_{[1,2]}^3 = \frac{1}{\sqrt 2} X_{ud\bar s}', \nonumber\\
 && X_{[1,2]}^1 = X_{[2,3]}^3 =  \frac{1}{2} Y'_{(u\bar u,s\bar s)d },\;\;\; X_{[3,1]}^1 = X_{[2,3]}^2 = \frac{1}{2} Y'_{(u\bar u, d\bar d)s},\;\;\;  X_{[1,2]}^2= X_{[3,1]}^3 = \frac{1}{2} Y'_{(d\bar d, s\bar s)u}.
\end{eqnarray}
The $ {\bf 15}$ representation is denoted  as $X^k_{\{i,j\}}$. Here the representation  is symmetric when exchanging i and j, and traceless $X^i_{\{i,j\}} = 0$ with the components~\cite{He:2016yhd}:
\begin{eqnarray}
&& X_{\{2,3\}}^1 = \frac{1}{\sqrt 2} X_{ds\bar u}, \;\;\; X_{\{3,1\}}^2 = \frac{1}{\sqrt 2} X_{su\bar d},\;\;\; X_{\{1,2\}}^3 = \frac{1}{\sqrt 2} X_{ud\bar s}, \nonumber\\
&& X_{\{1,1\}}^1 = \left( \frac{Y_{\pi u}}{\sqrt 2} + \frac{Y_{\eta u}}{\sqrt 6} \right),\;\;\; X_{\{1,2\}}^1 =  \frac{1}{\sqrt 2}  \left( \frac{Y_{\pi d}}{\sqrt 2} + \frac{Y_{\eta d}}{\sqrt 6} \right),\;\;\; X_{\{1,3\}}^1 =  \frac{1}{\sqrt 2} \left( \frac{Y_{\pi s}}{\sqrt 2} + \frac{Y_{\eta s}}{\sqrt 6} \right), \nonumber\\
&& X_{\{2,1\}}^2 =  \frac{1}{\sqrt 2}  \left( -\frac{Y_{\pi u}}{\sqrt 2} + \frac{Y_{\eta u}}{\sqrt 6} \right),\;\;\; X_{\{2,2\}}^2 =   \left( -\frac{Y_{\pi d}}{\sqrt 2} + \frac{Y_{\eta d}}{\sqrt 6} \right),\;\;\; X_{\{2,3\}}^2 =     \frac{1}{\sqrt 2}   \left(-\frac{Y_{\pi s}}{\sqrt 2} + \frac{Y_{\eta s}}{\sqrt 6} \right),\nonumber\\
&& X_{\{3,1\}}^3 = - \frac{Y_{\eta u}}{\sqrt 3},\;\;\; X_{\{3,2\}}^3 =    -\frac{Y_{\eta d}}{\sqrt 3},\;\;\; X_{\{3,3\}}^3 =   - \frac{Y_{\eta s}}{\sqrt 3},\nonumber\\
&& X_{\{2,2\}}^1 =Z_{dd\bar u} ,\;\;\; X_{\{3,3\}}^1 =Z_{ss\bar u} ,\nonumber\\
&& X_{\{1,1\}}^2 =Z_{uu\bar d} ,\;\;\; X_{\{3,3\}}^2 =Z_{ss\bar d} ,\nonumber\\
&& X_{\{1,1\}}^3 =Z_{uu\bar s} ,\;\;\; X_{\{2,2\}}^3 =Z_{dd\bar s}.
\end{eqnarray}

It is clear that if $SU(3)$ flavor symmetry plays an important  role in classifying charmed tetraquark states, there are associated members with those tetraquarks with three different light quarks. Whether they come as a ${\bf \bar 6}$ or ${\bf 15}$ has to be determined experimentally.

\subsection{Estimation of $X_c$ masses}

Before discussing $B_c(B) \to X_c P$, we estimate the masses for $X_c$.
Here we focus on the lowest-lying tetraquarks where their orbital
angular momenta are zero. We assume that the $X_c$ mass is from the constituent quark masses and also various spin-spin correlations as  proposed in Ref.~\cite{Maiani:2004vq,Maiani:2005pe}. This approach has been applied  to various multiquark systems~\cite{Ali:2009pi,Ali:2009es,Ali:2010pq,Ali:2011ug,Ali:2014dva,Zhu:2015bba,Maiani:2015vwa,Esposito:2016itg}. The effective Hamiltonian is   given by
\begin{eqnarray}
 H&=&m_{\delta}+m_{q''}+m_c+H^{\delta}_{SS} + H^{\bar{q}''\bar{c}}_{SS}+H^{\delta\bar{q}''}_{SS}+H^{\delta\bar{c}}_{SS} ,
\label{eq:definition-hamiltonian}
\end{eqnarray}
with the spin-spin interactions
\begin{eqnarray}
 &&H^\delta_{SS}=2(\kappa_{q q^\prime})_{\bar{3}}(\mathbf{S}_q\cdot \mathbf{S}_{q^\prime}),\nonumber\\
 &&H^{\bar{q}''\bar{c}}_{SS}=2(\kappa_{ cq''})_{\bar{3}}(\mathbf{S}_{\bar{c}}\cdot \mathbf{S}_{\bar{q}''}), \nonumber\\
 &&H^{\delta\bar{q}''}_{SS} =2\kappa_{q\bar{q}''}(\mathbf{S}_q\cdot \mathbf{S}_{\bar{q}''}) +2\kappa_{q^{\prime} \bar{q}''}(\mathbf{S}_{q^{\prime}}\cdot \mathbf{S}_{\bar{q}''}), \nonumber\\
 &&H^{\delta\bar{c}}_{SS} =2\kappa_{q\bar{c}} (\mathbf{S}_q\cdot \mathbf{S}_{\bar{c}})+ 2\kappa_{q^\prime\bar{c}}(\mathbf{S}_{q^\prime}\cdot \mathbf{S}_{\bar{c}})\ .
\label{eq:definition-hamiltonian2}
\end{eqnarray}
In the above, the $m_\delta$  is the constituent mass of the two quarks $[qq^\prime]$   to form a diquark $\delta$.  
The spin operator  of light quarks and  heavy antiquark  is  $\mathbf{S}_{q^{(\prime)}}$ and $\mathbf{S}_{\bar{c}}$, respectively. The spin-spin interaction inside the diquark is denoted as  $ H^\delta_{SS}$, while the  $ H^{\bar{q}''\bar{c}}_{SS}$ is the spin-spin interaction between two antiquarks. The $ H^{\delta\bar{q}''}_{SS}$ and $ H^{\delta\bar{c}}_{SS} $  reflect  the spin-spin interaction  between the quark and antiquark.   The orbital-related terms are neglected  for S-wave tetraquark states.  The coefficients $\kappa_{q_1\bar{q}_2}$ and $(\kappa_{q_1 q^\prime_2})_{\bar{3}}$ correspond  to the spin-spin coupling strengths.

The wave function of a tetraquark  consists of four parts, i.t. space-coordinate, color, flavor, and spin subspaces:
 \begin{eqnarray}
\Psi(q,q',\bar{q}'',\bar{c})&=&\psi(x_1,x_2,x_3,x_4)\otimes
\chi_c(c_1,c_2,c_3,c_4)\otimes \chi_f(f_1,f_2,f_3,f_4)\otimes \chi_s(s_1,s_2,s_3,s_4)\,,
\end{eqnarray}
where we use the labels 1, 2, 3, 4 to denote  $q$, $q'$, $\bar{q}''$, $\bar{c}$, respectively; $\psi(x_i)$, $\chi_c(c_i)$, $\chi_f(f_i)$, and $\chi_s(s_i)$ denote the space, color, flavor, and spin wave functions, respectively. 
Since we focus on the tetraquarks with $L=0$,
the space wave function is symmetric. The  diquark is attractive only in the triplet representation in color space, thus the color wave function is antisymmetric.

If the spin wave function of the light quark system $[qq']$ is antisymmetric, i.e. $S_{\delta}=0$, the flavor function should be also antisymmetric. In this case, the charmed  tetraquarks  can be decomposed into the ${\bf\bar 6}$ representation, with  the spin-parity $J^P=0^+,~1^+$:
\begin{eqnarray}
|0_\delta,\frac{1}{2}_{\bar{q}''};\frac{1}{2}_{\delta\bar{q}''},\frac{1}{2}_{\bar{c}},0_J\rangle&=&\frac{1}{2}
\big[(\uparrow)_q(\downarrow)_{q^\prime}-(\downarrow)_q(\uparrow)_{q^\prime} \big]\big[(\uparrow)_{\bar{q}''}(\downarrow)_{\bar{c}}
-(\downarrow)_{\bar{q}''}(\uparrow)_{\bar{c}}\big] ,\nonumber\\
|0_\delta,\frac{1}{2}_{\bar{q}''};\frac{1}{2}_{\delta\bar{q}''},\frac{1}{2}_{\bar{c}},1_J\rangle&=&\frac{1}{\sqrt{2}}
\big[(\uparrow)_q(\downarrow)_{q^\prime}-(\downarrow)_q(\uparrow)_{q^\prime} \big](\uparrow)_{\bar{q}''}(\uparrow)_{\bar{c}},
 \label{eq:definition-states0+}
\end{eqnarray}
In the above, the tetraquark states are classified according to  $|S_\delta,S_{\bar{q}''};S_{\delta\bar{q}''},S_{\bar{c}},S_J\rangle$; the $S_\delta$, $S_{\bar{q}''}$, $S_{\bar{c}}$ and $S_{\delta\bar{q}''}$ stand for  the spins  of diquark $[qq^\prime]$, antiquark, heavy antiquark, and $[qq']\bar{q}''$, respectively, while the $S_J$ is the total angular momentum.

Using the basis defined in Eq. (\ref{eq:definition-states0+}), one can derive  the mass matrix for the $J^P=0^+$ and $J^P=1^+$ tetraquarks in the ${\bf\bar 6}$ representation
\begin{eqnarray}
M(0^+)= m_{\delta}+m_{q''}+m_c-\frac{3}{2}\left((\kappa_{q{q^\prime}})_{\bar{3}}+(\kappa_{c q''})_{\bar{3}}\right)\,,
\nonumber\\
M(1^+)= m_{\delta}+m_{q''}+m_c-\frac{1}{2}\left(3(\kappa_{q{q^\prime}})_{\bar{3}}-(\kappa_{c q''})_{\bar{3}}\right)\,,
\end{eqnarray}
where we use the flavor SU(3) symmetry $\kappa_{q\bar{c}}=\kappa_{q'\bar{c}}$ and $\kappa_{q\bar{q}''}=\kappa_{q'\bar{q}''}$  for simplification, and this leads to the vanishing contribution from both $ H^{\delta\bar{q}''}_{SS}$ and $ H^{\delta\bar{c}}_{SS}$ interactions, because $\mathbf{S}_q+\mathbf{S}_{q'}=\mathbf{S}_\delta=0$.

If the spin wave function  of the light quark system $[qq']$ is symmetric, namely  $S_{\delta}=1$, the flavor function is also  symmetric. In this case, the charmed tetraquarks can be decomposed into the ${\bf 15}$ representation.
The spin-parity could be $J^P=0^+,~1^+,~2^+$: 
\begin{eqnarray}
|1_\delta,\frac{1}{2}_{\bar{q}''};\frac{1}{2}_{\delta\bar{q}''},\frac{1}{2}_{\bar{c}},0_J\rangle&=&\frac{1}{\sqrt{3}}
\big\{(\uparrow)_q(\uparrow)_{q^\prime}(\downarrow)_{\bar{q}''}(\downarrow)_{\bar{c}}
+(\downarrow)_q(\downarrow)_{q^\prime}(\uparrow)_{\bar{q}''}(\uparrow)_{\bar{c}} -\frac{1}{2}
\big[(\uparrow)_q(\downarrow)_{q^\prime}+(\downarrow)_q(\uparrow)_{q^\prime} \big]\big[(\uparrow)_{\bar{q}''}(\downarrow)_{\bar{c}}
+(\downarrow)_{\bar{q}''}(\uparrow)_{\bar{c}}\big]\big\},\nonumber\\
|1_\delta,\frac{1}{2}_{\bar{q}''};\frac{1}{2}_{\delta\bar{q}''},\frac{1}{2}_{\bar{c}},1_J\rangle&=&\frac{1}{\sqrt{6}}
\big\{2(\uparrow)_q(\uparrow)_{q^\prime}(\downarrow)_{\bar{q}''}(\uparrow)_{\bar{c}}
 -
\big[(\uparrow)_q(\downarrow)_{q^\prime}+(\downarrow)_q(\uparrow)_{q^\prime} \big](\uparrow)_{\bar{q}''}(\uparrow)_{\bar{c}}\big\},\nonumber\\
|1_\delta,\frac{1}{2}_{\bar{q}''};\frac{3}{2}_{\delta\bar{q}''},\frac{1}{2}_{\bar{c}},1_J\rangle&=&\frac{1}{2\sqrt{3}}
\big\{3(\uparrow)_q(\uparrow)_{q^\prime}(\uparrow)_{\bar{q}''}(\downarrow)_{\bar{c}}
 -(\uparrow)_q(\uparrow)_{q^\prime}(\downarrow)_{\bar{q}''}(\uparrow)_{\bar{c}}-
\big[(\uparrow)_q(\downarrow)_{q^\prime}+(\downarrow)_q(\uparrow)_{q^\prime} \big](\uparrow)_{\bar{q}''}(\uparrow)_{\bar{c}}\big\},\nonumber\\
|1_\delta,\frac{1}{2}_{\bar{q}''};\frac{3}{2}_{\delta\bar{q}''},\frac{1}{2}_{\bar{c}},2_J\rangle&=
&(\uparrow)_q(\uparrow)_{q^\prime}(\uparrow)_{\bar{q}}(\uparrow)_{\bar{Q}}.
 \label{eq:definition-states012+}
\end{eqnarray}

The masses for the $J^P=0^+$ and $J^P=2^+$ tetraquarks in the ${\bf15}$ representation are given as
\begin{eqnarray}
M(0^+)= m_{\delta}+m_{q''}+m_c+\frac{1}{2}\left((\kappa_{q{q^\prime}})_{\bar{3}}+(\kappa_{c q''})_{\bar{3}}\right)-2\left(\kappa_{q\bar{q''}}+\kappa_{q\bar{c}}\right)\,,
\nonumber\\
M(2^+)= m_{\delta}+m_{q''}+m_c+\kappa_{q\bar{q''}}+\kappa_{q\bar{c}}+\frac{1}{2}\left((\kappa_{q{q^\prime}})_{\bar{3}}+(\kappa_{c q''})_{\bar{3}}\right)\,.
\end{eqnarray}
Note that there are two possible ways for the charmed tetraquark with spin-parity  $J^P=1^+$.  One of them is from   the light quark system $qq' \bar q''$ having the spin $\frac{1}{2}$ and combing to the total spin 1 with the heavy antiquark, while the other one is from   the light quark system $qq' \bar q''$ having the spin $\frac{3}{2}$ and combing to the total spin 1 with the heavy antiquark. They mix with each other, since they have the same quantum numbers. 
Using the second and third basis   defined in Eq. (\ref{eq:definition-states012+}), one can obtain  the mass matrix $ M$ for $J^P=1^+$ tetraquarks in the ${\bf15}$ representation
\begin{eqnarray}
 M(1^+)=m_{\delta}+m_{q''}+m_c+\frac{1}{2}(\kappa_{q{q^\prime}})_{\bar{3}}+\left(
\begin{array}{cc}
 -2 \kappa_{q \bar{q}'}+\frac{2 }{3}\kappa_{q \bar{c}}-\frac{1}{6}(\kappa_{q''c})_{\bar{3}} & \frac{2}{3} \sqrt{2} ((\kappa_{q''c})_{\bar{3}}-\kappa_{q \bar{c}}) \\
 \frac{2}{3} \sqrt{2} ((\kappa_{q''c})_{\bar{3}}-\kappa_{q \bar{c}}) & \kappa_{q \bar{q}'}-\frac{5}{6} (2 \kappa_{q \bar{c}}+(\kappa_{q''c})_{\bar{3}})
\end{array}
\right).
\end{eqnarray}
Diagonalizing  the above matrix, one obtains two different eigenvalues of masses.

In the flavor SU(3) symmetry, all  charmed tetraquark states
will have the identical masses.  By distinguishing the strange quark from the up and down quarks, one can  obtain the charmed tetraquark masses including the SU(3) symmetry breaking effects. In the numerical  calculation, we will use the  quark masses as $m_q=305\mathrm{MeV}, m_s=490\mathrm{MeV},m_c=1.670\mathrm{GeV}$~\cite{Maiani:2004vq,Zhu:2015bba}.
For the light diquark $\delta=[qq]$, we use $m_{qq}=0.395${GeV} and $m_{sq}=0.590${GeV}~\cite{Maiani:2004vq}.
The strange diquark mass $m_{ss}=0.785${GeV} is estimated by the relation $m_{ss}-m_{sq}=m_{sq}-m_{qq}$.
The spin-spin couplings are $(\kappa _{qq})_{\bar 3}=103$MeV, $(\kappa _{sq})_{\bar 3}=64$MeV, $(\kappa _{cq})_{\bar 3}=22$MeV, $(\kappa _{cs})_{\bar 3}=25$MeV, $(\kappa _{ss})_{\bar 3}=72$MeV, $(\kappa _{q\bar{q}})_{0}=315$MeV, $(\kappa _{s\bar{q}})_{0}=195$MeV, $(\kappa _{s\bar{s}})_{0}=121$MeV, $(\kappa _{c\bar{q}})_{0}=70$MeV and $(\kappa _{c\bar{s}})_{0}=72$MeV~\cite{Maiani:2004vq,Zhu:2015bba}. The relation $\kappa _{ij}=\frac{1}{4}(\kappa _{ij})_{0}$ for the quark-antiquark state  derived from one gluon exchange model has been employed.

For the tetraquarks in  the ${\bf\bar 6}$  representation,  their masses are estimated to be:
\begin{align}
 m(X_{ds\bar u}')&= m(X_{su\bar d}')  = m(Y_{(u\bar u, d\bar d)s}')= \left\{ \begin{array} {ll}2.44 {\rm GeV}\ , & J^P=0^+ \ , \\ 2.48 {\rm GeV}\ , & J^P=1^+\ , \end{array} \right.\\
  m(X_{ud\bar s}') &= \left\{ \begin{array} {ll} 2.36 {\rm GeV}\ , & J^P=0^+ \ , \\ 2.41 {\rm GeV}\ , & J^P=1^+\ ,
  \end{array} \right.\\
   m(Y_{(u\bar u, s\bar s)d}')&=  m(Y_{(d\bar d, s\bar s)u}')  = \left\{ \begin{array} {ll}2.40 {\rm GeV}\ , & J^P=0^+ \ , \\ 2.45 {\rm GeV}\ , & J^P=1^+\ .
   \end{array} \right.
\end{align}
The spin of charmed tetraquark  states  in  the ${\bf15}$ representation could be 0, 1, and 2.  We give the predictions for  their masses:
\begin{align}
 m(X_{ds\bar u})&= m(X_{su\bar d})  = m(Y_{\pi s})= \left\{ \begin{array} {ll} 2.47 {\rm GeV}\ , & J^P=0^+ \ , \\
 2.51 {\rm GeV}, 2.60 {\rm GeV}\ , & J^P=1^+ \ , \\ 2.67 {\rm GeV}\ , & J^P=2^+\ , \end{array} \right.\\
  m(X_{ud\bar s}) &= m(Z_{uu\bar s})= m(Z_{dd\bar s})=\left\{ \begin{array} {ll} 2.49 {\rm GeV}\ , & J^P=0^+ \ , \\
 2.52 {\rm GeV}, 2.61 {\rm GeV}\ , & J^P=1^+ \ , \\ 2.69 {\rm GeV}\ , & J^P=2^+\ , \end{array} \right.\\
   m(Y_{\pi u}) &= m(Y_{\pi d})=m(Z_{uu\bar d})= m(Z_{dd\bar u})=\left\{ \begin{array} {ll} 2.24 {\rm GeV}\ , & J^P=0^+ \ , \\
 2.27 {\rm GeV}, 2.45 {\rm GeV}\ , & J^P=1^+ \ , \\ 2.53 {\rm GeV}\ , & J^P=2^+\ , \end{array} \right.\\
 m(Y_{\eta u}) &= m(Y_{\eta d})=\left\{ \begin{array} {ll} 2.55 {\rm GeV}\ , & J^P=0^+ \ , \\
 2.58 {\rm GeV}, 2.66 {\rm GeV}\ , & J^P=1^+ \ , \\ 2.74 {\rm GeV}\ , & J^P=2^+\ , \end{array} \right.\\
  m(Y_{\eta s}) &=\left\{ \begin{array} {ll} 2.76 {\rm GeV}\ , & J^P=0^+ \ , \\
 2.79 {\rm GeV}, 2.84 {\rm GeV}\ , & J^P=1^+ \ , \\ 2.92 {\rm GeV}\ , & J^P=2^+\ , \end{array} \right.\\
m(Z_{ss\bar u}) &= m(Z_{ss\bar d})=\left\{ \begin{array} {ll} 2.71 {\rm GeV}\ , & J^P=0^+ \ , \\
 2.74 {\rm GeV}, 2.78 {\rm GeV}\ , & J^P=1^+ \ , \\ 2.86 {\rm GeV}\ , & J^P=2^+\ . \end{array} \right.
\end{align}

The masses of $X_c$ are in the range between 2.24GeV and 2.92GeV.
The above discussion shows that there should be enough phase space to allow  $B_c(B) \to X_c P$ occur.

\section{Effective Hamiltonian and decay amplitudes for  $B_c(B) \to X_c P$}
\label{sec:effective_hamiltonian}

In this section, we study the $B_c \to X_c P$ and $B\to X_c P$  decays. Let us first identify the flavor $SU(3)$ symmetry properties of the particles involved. As already mentioned before  $X_c$ can be in ${\bf \bar 6}$ or ${\bf 15}$. The other  $SU(3)$ transformation properties are: the $B_c$ is a singlet in the SU(3), while the $B_i = (B_u(u\bar b), B_d(d \bar b), B_s(s \bar b))$ transform as ${\bf 3}$ representation, and the pseudo-scalar meson $P$ is an octet:
\begin{eqnarray}
 P^i_j=\begin{pmatrix}
 \frac{\pi^0}{\sqrt{2}}+\frac{\eta}{\sqrt{6}}  &\pi^+ & K^+\\
 \pi^-&-\frac{\pi^0}{\sqrt{2}}+\frac{\eta}{\sqrt{6}}&{K^0}\\
 K^-&\bar K^0 &-2\frac{\eta}{\sqrt{6}}
 \end{pmatrix}.
\end{eqnarray}

\subsection{$B_c \to X_c P$ decays}

The $B_c \to X_c P$ decays are induced by charmless $b\to q$ ($q=d,s$) transition.
The weak Hamiltonian ${\cal H}_{eff}$ is given by:
 \begin{eqnarray}
 {\cal H}_{eff} &=& \frac{G_{F}}{\sqrt{2}}
     \bigg\{ V_{ub} V_{uq}^{*} \big[
     C_{1}  O^{\bar uu}_{1}
  +  C_{2}  O^{\bar uu}_{2}\Big]- V_{tb} V_{tq}^{*} \big[{\sum\limits_{i=3}^{10}} C_{i}  O_{i} \Big]\bigg\}+ \mbox{H.c.} ,
 \label{eq:hamiltonian}
\end{eqnarray}
where the $V_{ij}$  is  CKM matrix element. The $O_{i}$ is a four-quark operator or a moment type operator. At the hadron level,  penguin operators  behave as the ${\bf\bar 3}$ representation while  tree operators in Eq.~\eqref{eq:hamiltonian} transform
under the flavor SU(3) symmetry as ${\bf\bar 3}\otimes {\bf3}\otimes {\bf\bar
3}={\bf\bar 3}\oplus {\bf\bar 3}\oplus {\bf6}\oplus {\bf\overline{15}}$. So the Hamiltonian  can
be decomposed in terms of a vector $H^i({\bf\overline3})$, a traceless
tensor antisymmetric in upper indices, $H^{[ij]}_k({\bf6})$, and a
traceless tensor symmetric in   upper indices,
$H^{\{ij\}}_k({\bf\overline{15}})$.

For the $\Delta S=0 (b\to d)$decays, the non-zero components of the effective Hamiltonian are~\cite{Savage:1989ub,He:2000ys,Hsiao:2015iiu}:
\begin{eqnarray}
 H^2({\bf\overline{3}})=1,\;\;\;H^{12}_1({\bf6})=-H^{21}_1({\bf6})=H^{23}_3({\bf6})=-H^{32}_3({\bf6})=1,\nonumber\\
 2H^{12}_1({\bf\overline{15}})= 2H^{21}_1({\bf\overline{15}})=-3H^{22}_2({\bf\overline{15}})=
 -6H^{23}_3({\bf\overline{15}})=-6H^{32}_3({\bf\overline{15}})=6,\label{eq:H3615}
\end{eqnarray}
with all other remaining entries zero. For the $\Delta S=1(b\to s)$
decays the nonzero entries in $H^i({\bf\overline{3}})$, $H^{[ij]}_k({\bf6})$,
$H^{(ij)}_k({\bf\overline{15}})$ are obtained from Eq.~\eqref{eq:H3615}
with the exchange  $2\leftrightarrow 3$.

\begin{table}
\caption{Decay amplitudes of $B_c \to X_c P$ decays into a ${\bf\bar 6}$ charmed tetraquark.  }\begin{tabular}{|c|c|c|c|}\hline\hline
channel $\Delta S=0$ & amplitude & channel $\Delta S = 1$ &amplitude \\\hline

$ B_c^-\to K^-  X'_{ud\bar s}$ & $ -\frac{a_3-2 a_6-a_{15}+b_6}{\sqrt{2}}$
&$ B_c^-\to \pi^-  X'_{su\bar d}$ &$ \frac{a_3-2 a_6-a_{15}+b_6}{\sqrt{2}}$\\\hline

$ B_c^-\to \pi^-  Y'_{(d\bar d, s\bar s)u}$ & $ -\frac{a_3-2 a_6-a_{15}+b_6}{2}$
&$ B_c^-\to K^-  Y'_{(d\bar d, s\bar s)u}$ &$ \frac{a_3-2 a_6-a_{15}+b_6}{2}$\\\hline

$ B_c^-\to K^+  X'_{ds\bar u}$ & $ \frac{a_3+2 a_6+3 a_{15}-b_6}{\sqrt{2}}$
&$ B_c^-\to \pi^+  X'_{ds\bar u}$ &$- \frac{a_3+2 a_6+3 a_{15}-b_6}{\sqrt{2}}$\\\hline

$ B_c^-\to K^0  Y'_{(u\bar u, d\bar d)s}$ & $ \frac{a_3+2 a_6-5 a_{15}-b_6}{2}$
&$ B_c^-\to \overline K^0  Y'_{(u\bar u, s\bar s)d}$ &$ -\frac{a_3+2 a_6-5 a_{15}-b_6}{2}$\\\hline

$ B_c^-\to \pi^0  Y'_{(u\bar u, s\bar s)d}$ & $- \frac{a_3-2 a_6+7 a_{15}+b_6}{2 \sqrt{2}}$
&$ B_c^-\to \pi^0  Y'_{(u\bar u, d\bar d)s}$ &$ \frac{a_3+a_{15}}{\sqrt{2}}$\\\hline

$ B_c^-\to \eta  Y'_{(u\bar u, s\bar s)d}$ & $ -\frac{3 a_3+2 a_6-3 a_{15}-b_6}{2 \sqrt{6}}$
&$ B_c^-\to \eta  Y'_{(u\bar u, d\bar d)s}$ &$ -\frac{2 a_6-6 a_{15}-b_6}{\sqrt{6}}$\\\hline

\end{tabular} \label{Bc2XcS016}
\end{table}

\begin{table}\caption{Decay amplitudes of $B_c\to X_c P$ decays into a tetraquark in the ${\bf 15}$ representation. }\begin{tabular}{|c|c|c|c|c|c|c|c}\hline\hline
channel $\Delta S=0$ & amplitude &channel $\Delta S=1$ & amplitude \\\hline

$ B_c^-\to K^-  X_{ud\bar s}$ &$ \frac{c_3+c_6+6 c_{15}-d_{15}}{\sqrt{2}}$
&$ B_c^-\to \pi^-  X_{su\bar d}$ &$ \frac{c_3+c_6+6 c_{15}-d_{15}}{\sqrt{2}}$\\\hline

$ B_c^-\to K^+  X_{ds\bar u}$ &$ \frac{c_3-c_6-2 c_{15}+3 d_{15}}{\sqrt{2}}$
&$ B_c^-\to \pi^+  X_{ds\bar u}$ &$ \frac{c_3-c_6-2 c_{15}+3 d_{15}}{\sqrt{2}}$\\\hline

$ B_c^-\to \pi^+  Z_{dd\bar u}$ &$ c_3-c_6-2 c_{15}+3 d_{15}$
&$ B_c^-\to K^+  Z_{ss\bar u}$ &$ c_3-c_6-2 c_{15}+3 d_{15}$\\\hline

$ B_c^-\to \overline K^0  Z_{dd\bar s}$ &$ c_3+c_6-2 c_{15}-d_{15}$
&$ B_c^-\to K^0  Z_{ss\bar d}$ &$ c_3+c_6-2 c_{15}-d_{15}$\\\hline

$ B_c^-\to \pi^0  Y_{\pi d}$ &$ \frac{\left(2+\sqrt{2}\right) c_3-2 \sqrt{2} c_6-2\left(2-3 \sqrt{2}\right) c_{15}-4 d_{15}}{4}$
&$ B_c^-\to \pi^0  Y_{\pi s}$ &$ \frac{c_3+2 c_{15}+d_{15}}{\sqrt{2}}$\\\hline

$ B_c^-\to \pi^0  Y_{\eta d}$ &$ \frac{\left(\sqrt{2}-2\right) c_3-4 \sqrt{2} c_6+2(3 \sqrt{2}+2) c_{15}-2( \sqrt{2}-2) d_{15}}{4 \sqrt{3}}$
&$ B_c^-\to \pi^0  Y_{\eta s}$ &$ \frac{-c_6+4 c_{15}+2 d_{15}}{\sqrt{6}}$\\\hline

$ B_c^-\to \pi^-  Y_{\pi u}$ &$ -\frac{c_3}{2}+\frac{c_6}{\sqrt{2}}-3 c_{15}+\frac{3 d_{15}}{\sqrt{2}}+d_{15}$
&$ B_c^-\to K^-  Y_{\pi u}$ &$ \frac{\left(1+\sqrt{2}\right) c_6+\left(1+3 \sqrt{2}\right) d_{15}}{2}$\\\hline

$ B_c^-\to \pi^-  Y_{\eta u}$ &$ \frac{c_3+\left(2+\sqrt{2}\right) c_6+6 c_{15}+3 \sqrt{2} d_{15}}{2 \sqrt{3}}$
&$ B_c^-\to \overline K^0  Y_{\pi d}$ &$ \frac{\left(1+\sqrt{2}\right) c_6+\left(3+\sqrt{2}\right) d_{15}}{2}$\\\hline

$ B_c^-\to K^0  Y_{\pi s}$ & $ -\frac{\left(c_3-c_6-2 c_{15}-5 d_{15}\right)}{2}$
&$ B_c^-\to \overline K^0  Y_{\eta d}$ &$ -\frac{2 c_3+\left(\sqrt{2}-1\right) c_6-4 c_{15}-(7-\sqrt{2}) d_{15}}{2 \sqrt{3}}$\\\hline

$ B_c^-\to K^0  Y_{\eta s}$ &$ \frac{c_3+3 c_6-2 c_{15}+3 d_{15}}{2 \sqrt{3}}$
&$ B_c^-\to K^-  Y_{\eta u}$ &$ -\frac{2 c_3-\left(\sqrt{2}-1\right) c_6-3 \left(-4 c_{15}+\left(1+\sqrt{2}\right) d_{15}\right)}{2 \sqrt{3}}$\\\hline

$ B_c^-\to \eta  Y_{\pi d}$ &$ \frac{\left(\sqrt{2}-2\right) c_3+2 \left(2+3 \sqrt{2}\right) \left(c_{15}+d_{15}\right)}{4 \sqrt{3}}$
&$ B_c^-\to \eta  Y_{\pi s}$ &$ -\frac{3 c_6-4 c_{15}+2 d_{15}}{\sqrt{6}}$\\\hline

$ B_c^-\to \eta  Y_{\eta d}$ &$ \frac{\left(2+5 \sqrt{2}\right) c_3+2 \left(3 \sqrt{2} c_6-\left(2+\sqrt{2}\right) c_{15}+2 \left(\sqrt{2}-1\right) d_{15}\right)}{12}$
&$ B_c^-\to \eta  Y_{\eta s}$ &$ \frac{3 c_3-2 c_{15}-5 d_{15}}{3 \sqrt{2}}$\\\hline

\end{tabular}\label{Bc2XcS0115}
\end{table}


The decay amplitudes $A(B_c\to X_c P) = \langle X_c P\vert {\cal H}_{eff} \vert B_c \rangle$ can be separated into two parts,   tree and the penguin
amplitudes $A^T_{B_c}$ and $A^P_{B_c}$, with
\begin{eqnarray}
A(B_c\to X_c P) = V_{ub}V^*_{uq}A^T_{B_c} + V_{tb}V^*_{tq}A^P_{B_c}\;.\label{BcA}
\end{eqnarray}
The $A^{\alpha= T, P}_{B_c}$ are then obtained by contracting the
$SU(3)$ indices in all possible ways. Each independent way will have an irreducible and non-perturbative  $SU(3)$ amplitude.
For the  ${\bf\bar 6}$ charmed tetraquark, we   have
\begin{eqnarray}
 A^\alpha_{B_c}= a^\alpha_3 H^i({\bf\bar 3}) X_{[i,j]}^k P^j_k
 + a^\alpha_6 H^{[ij]}_l ( {\bf6}) X_{[i,j]}^k P^l_k +  b^\alpha_6 H^{[il]}_k  ( {\bf6})X_{[i,j]}^k P^j_l+  a^\alpha_{15} H^{\{il\}}_k({\bf\overline {15}}) X_{[i,j]}^k P^j_l. \label{eq:amp_Bc_bar6}
\end{eqnarray}

For the  ${\bf15}$ charmed tetraquark, we  similarly have
\begin{eqnarray}
 A^\alpha_{B_c}= c^\alpha_3 H^i ({\bf\bar 3}) X_{\{i,j\}}^k P^j_k
 + c^\alpha_{15} H^{\{ij\}}_l ({\bf\overline {15}}) X_{\{i,j\}}^k P^l_k +  d^\alpha_{15} H^{\{il\}}_k ({\bf\overline {15}}) X_{\{i,j\}}^k P^j_l +   c^\alpha_{6} H^{[il]}_k({\bf6}) X_{\{i,j\}}^k P^j_l.\label{eq:amp_Bc_15}
\end{eqnarray}

Expanding the above expressions, one can  obtain the decay amplitudes $A^\alpha_{Bc}$. Results for the $A^T_{Bc}$ are collected
in Tables \ref{Bc2XcS016}, \ref{Bc2XcS0115}.
In these tables, we have dropped the superscript $\alpha=T$ for simplicity.  Penguin amplitudes $A^{P}_{Bc}$ have similar expressions but  only the triplet contributions containing  $\alpha_3^P$ are  nonzero.

\subsection{$B\to \overline{X}_c P$ decays}

\begin{table}
\caption{Decay amplitudes of $B \to \overline{X}_c P$ decays into a  charmed tetraquark in the ${\bf   6}$ representation. }
\begin{tabular}{|c|c|c|c|c|c|c|c}\hline\hline
channel $\Delta S=0$& amplitude &channel $\Delta S=1$ & amplitude\\\hline

$B^-\to K^-  \overline{X'}_{su\bar d}$ &$ -\frac{b_8+c_8}{\sqrt{2}}$
&$B^-\to \pi^-  \overline{X'}_{ud\bar s}$ &$ \frac{b_8+c_8}{\sqrt{2}}$\\\hline

$B^-\to \pi^-  \overline{Y'}_{(d\bar d, s\bar s)u}$ &$ \frac{b_8+c_8}{2}$
&$B^-\to K^-  \overline{Y'}_{(d\bar d, s\bar s)u}$ &$ -\frac{b_8+c_8}{2}$\\\hline

&&&\\\hline

$\overline B^0\to \overline K^0  \overline{X'}_{su\bar d}$ &$ -\frac{a_8+b_8}{\sqrt{2}}$
& $\overline B^0_s\to K^0  \overline{X'}_{ud\bar s}$ &$ \frac{a_8+b_8}{\sqrt{2}}$\\\hline

$\overline B^0\to \pi^-  \overline{Y'}_{(u\bar u, s\bar s)d}$ &$ \frac{a_8-d_8}{2}$
&$\overline B^0_s\to K^-  \overline{Y'}_{(u\bar u, d\bar d)s}$ &$ -\frac{a_8-d_8}{2}$\\\hline

$\overline B^0\to K^0  \overline{X'}_{ud\bar s}$ &$ \frac{a_8-d_8}{\sqrt{2}}$
&$\overline B^0_s\to \overline K^0  \overline{X'}_{su\bar d}$ &$- \frac{a_8-d_8}{\sqrt{2}}$\\\hline

$\overline B^0\to K^-  \overline{Y'}_{(u\bar u, d\bar d)s}$ &$- \frac{a_8-c_8}{2}$
&$\overline B^0_s\to \pi^-  \overline{Y'}_{(u\bar u, s\bar s)d}$ &$ \frac{a_8-c_8}{2}$\\\hline

$\overline B^0\to \pi^0  \overline{Y'}_{(d\bar d, s\bar s)u}$ &$ -\frac{a_8+b_8+c_8-d_8}{2 \sqrt{2}}$
&$\overline B^0_s\to \pi^0  \overline{Y'}_{(d\bar d, s\bar s)u}$ &$- \frac{a_8-c_8}{2 \sqrt{2}}$\\\hline

$\overline B^0\to \eta  \overline{Y'}_{(d\bar d, s\bar s)u}$ &$ \frac{3 a_8+b_8-c_8-d_8}{2 \sqrt{6}}$
&$\overline B^0_s\to \eta  \overline{Y'}_{(d\bar d, s\bar s)u}$ &$ \frac{3 a_8+2 b_8+c_8-2 d_8}{2 \sqrt{6}}$\\\hline

&&&\\\hline

$\overline B^0_s\to K^0  \overline{Y'}_{(d\bar d, s\bar s)u}$ &$ \frac{b_8+d_8}{2}$
&$\overline B^0\to \overline K^0  \overline{Y'}_{(d\bar d, s\bar s)u}$ &$ -\frac{b_8+d_8}{2}$\\\hline

$\overline B^0_s\to \pi^-  \overline{Y'}_{(u\bar u, d\bar d)s}$ &$ -\frac{c_8-d_8}{2}$
&$\overline B^0\to K^-  \overline{Y'}_{(u\bar u, s\bar s)d}$ &$ \frac{c_8-d_8}{2}$\\\hline

$\overline B^0_s\to \pi^0  \overline{X'}_{su\bar d}$ &$ \frac{c_8-d_8}{2}$
&$\overline B^0\to \pi^0  \overline{X'}_{ud\bar s}$ &$ -\frac{b_8+c_8}{2}$\\\hline

$\overline B^0_s\to \eta  \overline{X'}_{su\bar d}$ &$ \frac{2 b_8+c_8+d_8}{2 \sqrt{3}}$
&$\overline B^0\to \eta  \overline{X'}_{ud\bar s}$ &$ \frac{b_8-c_8+2 d_8}{2 \sqrt{3}}$\\\hline

\end{tabular} \label{B2aXcS016}
\end{table}


\begin{table}
\caption{Decay amplitudes of $B\to \overline{X}_c P$ decays into the ${\bf \overline{15}}$ representation.}\begin{tabular}{|c|c|c|c|c|c|c|c}\hline\hline
channel $\Delta S=0$ & amplitude &channel $\Delta S=1$ & amplitude \\\hline

$B^-\to K^-  \overline{X}_{su\bar d}$ &$ \frac{f_8+g_8}{\sqrt{2}}$
&$B^-\to \pi^-  \overline{X}_{ud\bar s}$ &$ \frac{f_8+g_8}{\sqrt{2}}$\\\hline

$B^-\to \pi^0  \overline{Z}_{uu\bar d}$ &$ \frac{f_8+g_8-h_8}{\sqrt{2}}$
&$B^-\to \pi^0  \overline{Z}_{uu\bar s}$ &$ \frac{f_8+g_8}{\sqrt{2}}$\\\hline

$B^-\to K^0  \overline{Z}_{uu\bar s}$ &$ h_8$
&$B^-\to \overline K^0  \overline{Z}_{uu\bar d}$ &$ h_8$\\\hline

$B^-\to \eta  \overline{Z}_{uu\bar d}$ &$ \frac{f_8+g_8+h_8}{\sqrt{6}}$
&$B^-\to \eta  \overline{Z}_{uu\bar s}$ &$ \frac{f_8+g_8-2 h_8}{\sqrt{6}}$\\\hline

$B^-\to \pi^-  \overline{Y}_{\pi u}$ &$ \frac{-f_8-g_8+\sqrt{2} h_8}{2}$
&$B^-\to K^-  \overline{Y}_{\pi u}$ &$ \frac{h_8}{\sqrt{2}}$\\\hline

$B^-\to \pi^-  \overline{Y}_{\eta u}$ &$ \frac{f_8+g_8+\sqrt{2} h_8}{2 \sqrt{3}}$
&$B^-\to K^-  \overline{Y}_{\eta u}$ &$ \frac{-2 f_8-2 g_8+\sqrt{2} h_8}{2 \sqrt{3}}$\\\hline

&&&\\\hline

$\overline B^0\to K^0  \overline{X}_{ud\bar s}$ &$ \frac{e_8+h_8}{\sqrt{2}}$
&$\overline B^0_s\to K^0  \overline{X}_{ud\bar s}$ &$ \frac{e_8+f_8}{\sqrt{2}}$\\\hline

$\overline B^0\to \overline K^0  \overline{X}_{su\bar d}$ &$ \frac{e_8+f_8}{\sqrt{2}}$
&$\overline B^0_s\to \overline K^0  \overline{X}_{su\bar d}$ &$ \frac{e_8+h_8}{\sqrt{2}}$\\\hline

$\overline B^0\to \pi^+  \overline{Z}_{uu\bar d}$ &$ e_8+f_8$
&$\overline B^0_s\to K^+  \overline{Z}_{uu\bar s}$ &$ e_8+f_8$\\\hline

$\overline B^0\to K^+  \overline{Z}_{uu\bar s}$ &$ e_8$
&$\overline B^0_s\to \pi^+  \overline{Z}_{uu\bar d}$ &$ e_8$\\\hline

$\overline B^0\to \pi^0  \overline{Y}_{\pi u}$ &$ \frac{\left(2+\sqrt{2}\right) e_8+\sqrt{2} \left(f_8-g_8+h_8\right)}{4}$
&$\overline B^0_s\to \pi^0  \overline{Y}_{\pi u}$ &$ \frac{\left(2+\sqrt{2}\right) e_8}{4}$\\\hline

$\overline B^0\to \pi^0  \overline{Y}_{\eta u}$ &$ -\frac{\left(\sqrt{2}-2\right) e_8+\sqrt{2} \left(f_8-g_8+h_8\right)}{4 \sqrt{3}}$
&$\overline B^0_s\to \pi^0  \overline{Y}_{\eta u}$ &$ \frac{-\sqrt{3} \left(\sqrt{2}-2\right) e_8-2 \sqrt{6} g_8}{12}$
\\\hline

$\overline B^0\to \pi^-  \overline{Y}_{\pi d}$ &$ \frac{e_8-\sqrt{2} g_8+h_8}{2}$
&$\overline B^0_s\to \pi^-  \overline{Y}_{\pi d}$ &$ \frac{e_8}{2}$\\\hline

$\overline B^0\to \pi^-  \overline{Y}_{\eta d}$ &$ \frac{e_8+\sqrt{2} g_8+h_8}{2 \sqrt{3}}$
&$\overline B^0_s\to \pi^-  \overline{Y}_{\eta d}$ &$ \frac{e_8-2 g_8}{2 \sqrt{3}}$\\\hline

$\overline B^0\to K^-  \overline{Y}_{\pi s}$ &$ \frac{e_8-g_8}{2}$
&$\overline B^0_s\to K^-  \overline{Y}_{\pi s}$ &$ \frac{e_8+h_8}{2}$\\\hline

$\overline B^0\to K^-  \overline{Y}_{\eta s}$ &$ \frac{e_8+g_8}{2 \sqrt{3}}$
&$\overline B^0_s\to K^-  \overline{Y}_{\eta s}$ &$ \frac{e_8-2 g_8+h_8}{2 \sqrt{3}}$\\\hline

$\overline B^0\to \eta  \overline{Y}_{\pi u}$ &$ -\frac{\left(\sqrt{2}-2\right) e_8+\sqrt{2} \left(f_8+g_8+h_8\right)}{4 \sqrt{3}}$
&$\overline B^0_s\to \eta  \overline{Y}_{\pi u}$ &$ -\frac{\left(\sqrt{2}-2\right) e_8}{4 \sqrt{3}}$\\\hline

$\overline B^0\to \eta  \overline{Y}_{\eta u}$ &$ \frac{\left(2+5 \sqrt{2}\right) e_8+\sqrt{2} \left(f_8+g_8+h_8\right)}{12}$
&$\overline B^0_s\to \eta  \overline{Y}_{\eta u}$ &$ \frac{\left(2+5 \sqrt{2}\right) e_8+2 \sqrt{2} \left(2 f_8-g_8+2 h_8\right)}{12}$\\\hline

&&&\\\hline

$\overline B^0_s\to \pi^0  \overline{X}_{su\bar d}$ &$ \frac{g_8-h_8}{2}$
& $\overline B^0\to \pi^0  \overline{X}_{ud\bar s}$ &$ \frac{g_8-f_8}{2}$ \\\hline

$\overline B^0_s\to \eta  \overline{X}_{su\bar d}$ &$ \frac{-2 f_8+g_8+h_8}{2 \sqrt{3}}$
&$\overline B^0\to \eta  \overline{X}_{ud\bar s}$ &$ \frac{f_8+g_8-2 h_8}{2 \sqrt{3}}$\\\hline

$\overline B^0_s\to K^+  \overline{Z}_{uu\bar d}$ &$ f_8$
&$\overline B^0\to \pi^+  \overline{Z}_{uu\bar s}$ &$ f_8$\\\hline

$\overline B^0_s\to K^-  \overline{Z}_{ss\bar d}$ &$ g_8$
&$\overline B^0\to \pi^-  \overline{Z}_{dd\bar s}$ &$ g_8$\\\hline

$\overline B^0_s\to K^0  \overline{Y}_{\pi u}$ &$ -\frac{f_8}{2}$
&$\overline B^0\to \overline K^0  \overline{Y}_{\pi u}$ &$ -\frac{h_8}{2}$\\\hline

$\overline B^0_s\to K^0  \overline{Y}_{\eta u}$ &$ \frac{f_8-2 h_8}{2 \sqrt{3}}$
&$\overline B^0\to \overline K^0  \overline{Y}_{\eta u}$ &$ \frac{h_8-2 f_8}{2 \sqrt{3}}$\\\hline

$\overline B^0_s\to \pi^-  \overline{Y}_{\pi s}$ &$ \frac{h_8-g_8}{2}$
&$\overline B^0\to K^-  \overline{Y}_{\pi d}$ &$ \frac{h_8}{2}$\\\hline

$\overline B^0_s\to \pi^-  \overline{Y}_{\eta s}$ &$ \frac{g_8+h_8}{2 \sqrt{3}}$
&$\overline B^0\to K^-  \overline{Y}_{\eta d}$ &$ \frac{h_8-2 g_8}{2 \sqrt{3}}$\\\hline

\end{tabular}\label{B2aXcS0115}
\end{table}


For $B \to \overline{X}_c(X_c) P$ decays, the $b$-quark in $B$ should decay into a charm (anti-charm) quark so that $X_c$ or its charge conjugate  can be generated.
The operator to produce a charm quark  from a $b$-quark, $\bar c b \bar q u$ is given by
\begin{eqnarray}
{\cal H}_{eff} &=& \frac{G_{F}}{\sqrt{2}}
     V_{cb} V_{uq}^{*} \big[
     C_{1}  O^{\bar cu}_{1}
  +  C_{2}  O^{\bar cu}_{2}\Big] +{\rm H.c.} . 
\end{eqnarray}
The light quarks in this effective Hamiltonian form an octet with the nonzero entry
\begin{eqnarray}
H^2_1({\bf8})=1
\end{eqnarray}
for   the $b\to c\bar ud$ transition, and   $H^3_1({\bf8})=1$ for  the $b\to c\bar
us$ transition.

Similarly as for the $B$ decays, one can write down the irreducible $SU(3)$ amplitudes for $B\to \overline{X}_c P$ decays.
This time, there exist only tree amplitudes which one can  normalize $A(B\to \overline{X}_c P) = V_{cb}V^*_{uq}A_{B\bar X_c}$.
For ${\bf 6}$ charmed tetraquarks, the decay amplitudes can be constructed as
\begin{eqnarray}
 A_{B\overline{X_c}}= a_8 B_i H^i_j({\bf8})  \overline{X}^{[j,l]}_k P^k_l  +b_8 B_i H^k_j({\bf8})  \overline{X}^{[j,l]}_k P^i_l  +c_8 B_j H^k_i({\bf8})  \overline{X}^{[j,l]}_k P^i_l      +d_8 B_j H^i_l ({\bf8}) \overline{X}^{[j,l]}_k P^k_i.\label{eq:amp_B_barXc_6}
\end{eqnarray}
For the  $  {\bf \overline{15}}$ representation, the effective Hamiltonian  is similar:
\begin{eqnarray}
 A_{B\overline{X_c}}= e_8 B_i H^i_j({\bf8})  \overline{X}^{\{j,l\}}_k P^k_l  +f_8 B_i H^k_j({\bf8})  \overline{X}^{\{j,l\}}_k P^i_l  +g_8 B_j H^k_i({\bf8})  \overline{X}^{\{j,l\}}_k P^i_l  +h_8 B_j H^i_l ({\bf8}) \overline{X}^{\{j,l\}}_k P^k_i.
\end{eqnarray}

Results for $A_{B\overline{X}_c}$ are obtained by expanding the above expressions and  summarized 
in Tables \ref{B2aXcS016} and \ref{B2aXcS0115}.

\subsection{$B\to X_c P$ decays}

For the anti-charm production, the operator having the quark contents $(\bar ub)(\bar qc)$  is given by
\begin{eqnarray}
{\cal H}_{eff} &=& \frac{G_{F}}{\sqrt{2}}
     V_{ub} V_{cq}^{*} \big[
     C_{1}  O^{\bar uc}_{1}
  +  C_{2}  O^{\bar uc}_{2}\Big]+ {\rm H.c.}. 
\end{eqnarray}
The two light anti-quarks form the ${\bf 3}$ and ${\bf \bar 6}$ representations.
The anti-symmetric tensor $H({\bf3})$ and the symmetric tensor
$H({\bf\bar 6})$ have nonzero components
\begin{eqnarray}
 H^{13}({\bf3})=-H^{31}({\bf3})=1,\;\;\; H^{13}({\bf\bar 6})=H^{31}({\bf\bar 6})=1,
\end{eqnarray}
for the $b\to u\bar cs$ transition. For the transition $b\to
u\bar cd$ one requests the interchange of $2\leftrightarrow 3$ in the
subscripts. The first kind of decays is proportional to
$|V_{cb}V_{ud}^*|\sim A\lambda^2$ while the second of decays is
proportional to $|V_{cd}V_{ub}^*|\sim A\lambda^4$, thus the latter
is greatly  suppressed compared to the first ones.


\begin{table}
\caption{Decay amplitudes of $B \to X_c P$ decays into the ${\bf \bar 6}$.}\begin{tabular}{|c|c|c|c|c|c|c|c}\hline\hline
channel $\Delta S=0$ & amplitude &channel $\Delta S=1$ & amplitude \\\hline

$B^-\to K^+  X'_{ds\bar u}$ &$ \frac{a_6+b_3+b_6+c_3}{\sqrt{2}}$
&$B^-\to \pi^+  X'_{ds\bar u}$ &$ -\frac{a_6+b_3+b_6+c_3}{\sqrt{2}}$\\\hline

$B^-\to K^-  X'_{ud\bar s}$ &$ \frac{2 a_3-a_6-b_3}{\sqrt{2}}$
&$B^-\to \pi^-  X'_{su\bar d}$ &$ \frac{-2 a_3+a_6+b_3}{\sqrt{2}}$\\\hline

$B^-\to \pi^-  Y'_{(d\bar d, s\bar s)u}$ &$ a_3-\frac{a_6}{2}-\frac{b_3}{2}$
&$B^-\to K^-  Y'_{(d\bar d, s\bar s)u}$ &$ \frac{-2 a_3+a_6+b_3}{2}$\\\hline

$B^-\to K^0  Y'_{(u\bar u, d\bar d)s}$ &$ \frac{a_6+b_3-b_6+c_3}{2}$
&$B^-\to \overline K^0  Y'_{(u\bar u, s\bar s)d}$ &$ \frac{-a_6-b_3+b_6-c_3}{2}$\\\hline

$B^-\to \pi^0  Y'_{(u\bar u, s\bar s)d}$ &$ \frac{2 a_3-a_6-b_3-2 b_6}{2 \sqrt{2}}$
&$B^-\to \pi^0  Y'_{(u\bar u, d\bar d)s}$ &$ \frac{-2 a_3+2 a_6+2 b_3+b_6+c_3}{2 \sqrt{2}}$\\\hline

$B^-\to \eta  Y'_{(u\bar u, s\bar s)d}$ &$ \frac{2 a_3-3 a_6-3 b_3-2 c_3}{2 \sqrt{6}}$
&$B^-\to \eta  Y'_{(u\bar u, d\bar d)s}$ &$ -\frac{2 a_3-3 b_6+c_3}{2 \sqrt{6}}$\\\hline

&&&\\\hline

$\overline B^0\to K^0  X'_{su\bar d}$ &$ \frac{-a_6+b_3-b_6+c_3}{\sqrt{2}}$
&$\overline B^0_s\to \overline K^0  X'_{ud\bar s}$ &$ \frac{a_6-b_3+b_6-c_3}{\sqrt{2}}$\\\hline

$\overline B^0\to \overline K^0  X'_{ud\bar s}$ &$ \frac{2 a_3+a_6-b_3}{\sqrt{2}}$
&$\overline B^0_s\to K^0  X'_{su\bar d}$ &$ \frac{-2 a_3-a_6+b_3}{\sqrt{2}}$\\\hline

$\overline B^0\to \pi^+  Y'_{(u\bar u, s\bar s)d}$ &$ \frac{2 a_3+a_6-b_3}{2}$
&$\overline B^0_s\to K^+  Y'_{(u\bar u, d\bar d)s}$ &$ \frac{-2 a_3-a_6+b_3}{2}$\\\hline

$\overline B^0\to K^+  Y'_{(u\bar u, d\bar d)s}$ &$ \frac{-a_6+b_3+b_6+c_3}{2}$
&$\overline B^0_s\to \pi^+  Y'_{(u\bar u, s\bar s)d}$ &$ \frac{a_6-b_3-b_6-c_3}{2}$\\\hline

$\overline B^0\to \pi^0  Y'_{(d\bar d, s\bar s)u}$ &$ -\frac{2 a_3+a_6-b_3+2 b_6}{2 \sqrt{2}}$
&$\overline B^0_s\to \pi^0  Y'_{(d\bar d, s\bar s)u}$ &$ \frac{-a_6+b_3+b_6+c_3}{2 \sqrt{2}}$\\\hline

$\overline B^0\to \eta  Y'_{(d\bar d, s\bar s)u}$ &$ \frac{2 a_3+3 a_6-3 b_3-2 c_3}{2 \sqrt{6}}$
&$\overline B^0_s\to \eta  Y'_{(d\bar d, s\bar s)u}$ &$ \frac{4 a_3+3 a_6-3 b_3+3 b_6-c_3}{2 \sqrt{6}}$\\\hline

&&&\\\hline

$\overline B^0_s\to K^0  Y'_{(d\bar d, s\bar s)u}$ &$ \frac{2 a_3-b_6+c_3}{2}$
&$\overline B^0\to \overline K^0  Y'_{(d\bar d, s\bar s)u}$ &$ \frac{-2 a_3+b_6-c_3}{2}$\\\hline

$\overline B^0_s\to K^+  Y'_{(u\bar u, s\bar s)d}$ &$ \frac{2 a_3+b_6+c_3}{2}$
&$\overline B^0\to \pi^+  Y'_{(u\bar u, d\bar d)s}$ &$ \frac{-2 a_3-b_6-c_3}{2}$\\\hline

$\overline B^0_s\to \pi^0  X'_{ud\bar s}$ &$ -b_6$
&$\overline B^0\to \pi^0  X'_{su\bar d}$ &$ \frac{2 a_3+b_6+c_3}{2}$\\\hline

$\overline B^0_s\to \eta  X'_{ud\bar s}$ &$ -\frac{2 a_3+c_3}{\sqrt{3}}$
&$\overline B^0\to \eta  X'_{su\bar d}$ &$ -\frac{2 a_3-3 b_6+c_3}{2 \sqrt{3}}$\\\hline

\end{tabular}\label{B2XcS016}
\end{table}

\begin{table}
\caption{Decay amplitudes of $B\to X_c P$ decays into the ${\bf 15}$ representation.}
\begin{tabular}{|c|c|c|c|c|c|c|c}\hline\hline
channel $\Delta S=0$ & amplitude &channel $\Delta S=1$ & amplitude  \\\hline

$B^-\to K^+  X_{ds\bar u}$ &$ \frac{d_3+e_3+e_6+f_6}{\sqrt{2}}$
&$B^-\to \pi^+  X_{ds\bar u}$ &$ \frac{d_3+e_3+e_6+f_6}{\sqrt{2}}$\\\hline

$B^-\to K^-  X_{ud\bar s}$ &$ \frac{d_3+2 d_6+e_6}{\sqrt{2}}$
&$B^-\to \pi^-  X_{su\bar d}$ &$ \frac{d_3+2 d_6+e_6}{\sqrt{2}}$\\\hline

$B^-\to \pi^+  Z_{dd\bar u}$ &$ d_3+e_3+e_6+f_6$
&$B^-\to K^+  Z_{ss\bar u}$ &$ d_3+e_3+e_6+f_6$\\\hline

$B^-\to \overline K^0  Z_{dd\bar s}$ &$ d_3+e_6$
&$B^-\to K^0  Z_{ss\bar d}$ &$ d_3+e_6$\\\hline

$B^-\to \pi^-  Y_{\pi u}$ &$ \frac{-d_3-2 d_6-\sqrt{2} e_3-e_6+\sqrt{2} f_6}{2}$
&$B^-\to K^-  Y_{\pi u}$ &$ \frac{f_6-e_3}{\sqrt{2}}$\\\hline

$B^-\to \pi^-  Y_{\eta u}$ &$ \frac{d_3+2 d_6-\sqrt{2} e_3+e_6+\sqrt{2} f_6}{2 \sqrt{3}}$
&$B^-\to K^-  Y_{\eta u}$ &$ -\frac{2 d_3+4 d_6+\sqrt{2} e_3+2 e_6-\sqrt{2} f_6}{2 \sqrt{3}}$\\\hline

$B^-\to \pi^0  Y_{\pi d}$ &$ \frac{\left(2+\sqrt{2}\right) d_3+2 \sqrt{2} d_6+2 \sqrt{2} e_3+\sqrt{2} e_6+2 e_6}{4}$
&$B^-\to \pi^0  Y_{\pi s}$ &$ \frac{2 d_3+2 d_6+e_3+2 e_6+f_6}{2 \sqrt{2}}$
\\\hline

$B^-\to \pi^0  Y_{\eta d}$ &$ \frac{\left(\sqrt{2}-2\right) d_3+2 \sqrt{2} d_6+2 \sqrt{2} e_3+\sqrt{2} e_6-2 e_6}{4 \sqrt{3}}$
&$B^-\to \pi^0  Y_{\eta s}$ &$ \frac{2 d_6+e_3+f_6}{2 \sqrt{6}}$\\\hline

$B^-\to K^0  Y_{\pi s}$ &$ \frac{-d_3-e_3-e_6+f_6}{2}$
&$B^-\to \overline K^0  Y_{\pi d}$ &$ \frac{f_6-e_3}{2}$\\\hline

$B^-\to K^0  Y_{\eta s}$ &$ \frac{d_3-e_3+e_6+f_6}{2 \sqrt{3}}$
&$B^-\to \overline K^0  Y_{\eta d}$ &$ -\frac{2 d_3+e_3+2 e_6-f_6}{2 \sqrt{3}}$\\\hline

$B^-\to \eta  Y_{\pi d}$ &$ \frac{\left(\sqrt{2}-2\right) d_3+2 \sqrt{2} d_6+\sqrt{2} e_6-2 e_6+2 \sqrt{2} f_6}{4 \sqrt{3}}$
&$B^-\to \eta  Y_{\pi s}$ &$ \frac{2 d_6+3 e_3-f_6}{2 \sqrt{6}}$\\\hline

$B^-\to \eta  Y_{\eta d}$ &$ \frac{\left(2+5 \sqrt{2}\right) d_3+2 \sqrt{2} d_6+5 \sqrt{2} e_6+2 e_6+2 \sqrt{2} f_6}{12}$
&$B^-\to \eta  Y_{\eta s}$ &$ \frac{6 d_3+2 d_6+3 e_3+6 e_6-f_6}{6 \sqrt{2}}$\\\hline

&&&\\\hline

$\overline B^0\to K^0  X_{su\bar d}$ &$ \frac{-d_3-e_3+e_6+f_6}{\sqrt{2}}$
&$\overline B^0_s\to \overline K^0  X_{ud\bar s}$ &$ \frac{-d_3-e_3+e_6+f_6}{\sqrt{2}}$\\\hline

$\overline B^0\to \overline K^0  X_{ud\bar s}$ &$ \frac{-d_3+2 d_6+e_6}{\sqrt{2}}$
&$\overline B^0_s\to K^0  X_{su\bar d}$ &$ \frac{-d_3+2 d_6+e_6}{\sqrt{2}}$\\\hline

$\overline B^0\to \pi^-  Z_{uu\bar d}$ &$ -d_3-e_3+e_6+f_6$
&$\overline B^0_s\to K^-  Z_{uu\bar s}$ &$ -d_3-e_3+e_6+f_6$\\\hline

$\overline B^0\to K^-  Z_{uu\bar s}$ &$ e_6-d_3$
&$\overline B^0_s\to \pi^-  Z_{uu\bar d}$ &$ e_6-d_3$\\\hline

$\overline B^0\to \pi^+  Y_{\pi d}$ &$ \frac{-d_3+2 d_6-\sqrt{2} e_3+e_6-\sqrt{2} f_6}{2}$
&$\overline B^0_s\to K^+  Y_{\pi s}$ &$ \frac{-d_3+2 d_6+e_6}{2}$\\\hline

$\overline B^0\to \pi^+  Y_{\eta d}$ &$ \frac{-d_3+2 d_6+\sqrt{2} e_3+e_6+\sqrt{2} f_6}{2 \sqrt{3}}$
&$\overline B^0_s\to K^+  Y_{\eta s}$ &$ -\frac{d_3-2 d_6+2 e_3-e_6+2 f_6}{2 \sqrt{3}}$\\\hline

$\overline B^0\to \pi^0  Y_{\pi u}$ &$ \frac{-\left(2+\sqrt{2}\right) d_3+2 \sqrt{2} d_6-2 \sqrt{2} e_3+\sqrt{2} e_6+2 e_6}{4}$
&$\overline B^0_s\to \pi^0  Y_{\pi u}$ &$ -\frac{\left (2+\sqrt{2}\right) \left(d_3-e_6\right )}{4}$\\\hline

$\overline B^0\to \pi^0  Y_{\eta u}$ &$ \frac{\left(\sqrt{2}-2\right) d_3-2 \sqrt{2} d_6+2 \sqrt{2} e_3-\sqrt{2} e_6+2 e_6}{4 \sqrt{3}}$
&$\overline B^0_s\to \pi^0  Y_{\eta u}$ &$ \frac{\left(\sqrt{2}-2\right) d_3-2 \sqrt{2} e_3-\sqrt{2} e_6+2 e_6-2 \sqrt{2} f_6}{4 \sqrt{3}}$\\\hline

$\overline B^0\to K^+  Y_{\pi s}$ &$ \frac{-d_3-e_3+e_6-f_6}{2}$
&$\overline B^0_s\to \pi^+  Y_{\pi d}$ &$ \frac{e_6-d_3}{2}$\\\hline

$\overline B^0\to K^+  Y_{\eta s}$ &$ \frac{-d_3+e_3+e_6+f_6}{2 \sqrt{3}}$
&$\overline B^0_s\to \pi^+  Y_{\eta d}$ &$ -\frac{d_3+2 e_3-e_6+2 f_6}{2 \sqrt{3}}$\\\hline

$\overline B^0\to \eta  Y_{\pi u}$ &$ \frac{\left(\sqrt{2}-2\right) d_3-2 \sqrt{2} d_6-\sqrt{2} e_6+2 e_6-2 \sqrt{2} f_6}{4 \sqrt{3}}$
&$\overline B^0_s\to \eta  Y_{\pi u}$ &$ \frac{\left(\sqrt{2}-2\right) \left(d_3-e_6\right)}{4 \sqrt{3}}$\\\hline

$\overline B^0\to \eta  Y_{\eta u}$ &$ \frac{-\left(2+5 \sqrt{2}\right) d_3+2 \sqrt{2} d_6+5 \sqrt{2} e_6+2 e_6+2 \sqrt{2} f_6}{12}$
&$\overline B^0_s\to \eta  Y_{\eta u}$ &$ \frac{-\left(2+5 \sqrt{2}\right) d_3+8 \sqrt{2} d_6-6 \sqrt{2} e_3+5 \sqrt{2} e_6+2 e_6+2 \sqrt{2} f_6}{12}$\\\hline

&&&\\\hline

$\overline B^0_s\to \pi^0  X_{ud\bar s}$ &$ e_3$
&$\overline B^0\to \pi^0  X_{su\bar d}$ &$ \frac{-2 d_6+e_3+f_6}{2}$\\\hline

$\overline B^0_s\to \eta  X_{ud\bar s}$ &$ \frac{f_6-2 d_6}{\sqrt{3}}$
&$\overline B^0\to \eta  X_{su\bar d}$ &$ \frac{2 d_6+3 e_3-f_6}{2 \sqrt{3}}$\\\hline

$\overline B^0_s\to \pi^+  Z_{dd\bar s}$ &$ e_3+f_6$
&$\overline B^0\to K^+  Z_{ss\bar d}$ &$ e_3+f_6$\\\hline

$\overline B^0_s\to \pi^-  Z_{uu\bar s}$ &$ f_6-e_3$
&$\overline B^0\to K^-  Z_{uu\bar d}$ &$ f_6-e_3$\\\hline

$\overline B^0_s\to K^+  Y_{\pi d}$ &$ d_6$
&$\overline B^0\to \pi^+  Y_{\pi s}$ &$ d_6-\frac{e_3}{2}-\frac{f_6}{2}$\\\hline

$\overline B^0_s\to K^+  Y_{\eta d}$ &$ \frac{d_6-e_3-f_6}{\sqrt{3}}$
&$\overline B^0\to \pi^+  Y_{\eta s}$ &$ \frac{2 d_6+e_3+f_6}{2 \sqrt{3}}$\\\hline

$\overline B^0_s\to K^0  Y_{\pi u}$ &$ -d_6$
&$\overline B^0\to \overline K^0  Y_{\pi u}$ &$ \frac{e_3-f_6}{2}$\\\hline

$\overline B^0_s\to K^0  Y_{\eta u}$ &$ \frac{d_6+e_3-f_6}{\sqrt{3}}$
&$\overline B^0\to \overline K^0  Y_{\eta u}$ &$ -\frac{4 d_6+e_3-f_6}{2 \sqrt{3}}$\\\hline

\end{tabular}\label{B2XcS0115}
\end{table}

Charmed tetraquarks can be produced by the $b\to u\bar cd(s)$ transition. The irreducible $SU(3)$ amplitudes can be obtained by repeating previous procedure for $B \to \overline{ X}_c P$ decays. We normalize $A(B\to X_c P) = V_{ub}V^*_{cq}A_{BX_c}$.
 For the  ${\bf\bar 6}$ charmed tetraquarks, the effective Hamiltonian can be constructed as
\begin{eqnarray}
 A_{BX_c} &=&
 a_3 B_i H^{[jl]}({\bf3})  X_{[j,l]}^k P^i_k  +b_3 B_i H^{[ij]}({\bf3})  X_{[j,l]}^k P^l_k   +c_3 B_k H^{[ij]}({\bf3})  X_{[j,l]}^k P^l_i  \nonumber\\
 &&  +a_{6} B_i H^{\{ij\}}({\bf6})  X_{[j,l]}^k P^l_k   +b_6 B_k H^{\{ij\}}({\bf6})  X_{[j,l]}^k P^l_i,
\end{eqnarray}
with the similar expression for the  $  {\bf15}$ representation:
\begin{eqnarray}
 A_{BX_c} &=&
 d_{3} B_i H^{[ij]}({\bf3})  X_{\{j,l\}}^k P^l_k   +e_{3} B_k H^{[ij]}({\bf3})  X_{\{j,l\}}^k P^l_i   \nonumber\\
 && + d_{6} B_i H^{\{jl\}}({\bf6})  X_{\{j,l\}}^k P^i_k  +e_{6} B_i H^{\{ij\}}({\bf6})  X_{\{j,l\}}^k P^l_k   +f_{6} B_j H^{\{ij\}} ({\bf6}) X_{\{j,l\}}^k P^l_i.
\end{eqnarray}

The results for  $A_{BX_c}$ are collected
in Tables \ref{B2XcS016}, and \ref{B2XcS0115}.

\section{Results and Discussions: Decay Widths and CP Violation}
\label{sec:discussions}

Expanding the irreducible $SU(3)$ amplitudes obtained in the previous section, one obtains the decay amplitudes for each individual decays. They are listed in Tables \ref{Bc2XcS016} to \ref{B2XcS0115}.  Due to the non-perturbative nature of QCD, we are far from being able to calculate these amplitudes from first principle. We will not attempt to do  so in this work. Instead, we will use the amplitudes obtained to derive  relations which may be testable  experimentally.
We find three types of relations of interest  which related two different decays. We provide some details in the following.

\subsection{Equal or proportional decay widths with known coefficients}

\begin{table}
\caption{SU(3) relations for decay widths for the ${\bf 6}$(${\bf \bar{6}}$)  and ${\bf 15}$(${\bf \overline{15}}$). Herein and in the following tables, the left three columns are for the ${\bf 15}$ or ${\bf \overline{15}}$; while the right three columns are for the ${\bf 6}$ or ${\bf \bar{6}}$.
 $R$ denotes the ratio of two decay widths.} \label{decay-rate}
 \begin{tabular}{|c|c|c||c|c|c|}\hline\hline
channel 1 & channel 2 & $R$  & channel 1 & channel 2 & $R$ \\\hline
$B_c^-\to \pi^+  Z_{dd\bar u}$ & $B_c^-\to K^+  X_{ds\bar u}$ & $2$  &$B_c^-\to K^-  X'_{ud\bar s}$ & $B_c^-\to \pi^-  Y'_{(d\bar d, s\bar s)u}$ &  $2$\\ \hline\\\hline
$B_c^-\to K^+  Z_{ss\bar u}$ & $B_c^-\to \pi^+  X_{ds\bar u}$ & $2$    &$B_c^-\to \pi^-  X'_{su\bar d}$ & $B_c^-\to K^-  Y'_{(d\bar d, s\bar s)u}$ & $2$\\ \hline\\\hline
$\overline B^0\to \pi^+  \overline{Z}_{uu\bar d} $&$ \overline B^0\to \overline K^0  \overline{X}_{su\bar d}$&$2$ &$B^-\to K^-  \overline{X'}_{su\bar d} $&$ B^-\to \pi^-  \overline{Y'}_{(d\bar d, s\bar s)u}$ &$2$\\ \hline
$\overline B^0_s\to \pi^0  \overline{X}_{su\bar d} $&$ \overline B^0_s\to \pi^-  \overline{Y}_{\pi s}$ &$1$       &$\overline B^0\to K^0  \overline{X'}_{ud\bar s} $&$ \overline B^0\to \pi^-  \overline{Y'}_{(u\bar u, s\bar s)d}$ &$2$\\\hline
$\overline B^0_s\to K^+  \overline{Z}_{uu\bar d} $&$ \overline B^0_s\to K^0  \overline{Y}_{\pi u}$ &$4$           &$\overline B^0_s\to \pi^0  \overline{X'}_{su\bar d} $&$ \overline B^0_s\to \pi^-  \overline{Y'}_{(u\bar u, d\bar d)s}$&$1$\\ \hline\\\hline
$B^-\to \pi^0  \overline{Z}_{uu\bar s} $&$ B^-\to \pi^-  \overline{X}_{ud\bar s}$ &$1$                            &$B^-\to \pi^-  \overline{X'}_{ud\bar s} $&$ B^-\to K^-  \overline{Y'}_{(d\bar d, s\bar s)u}$ &$2$\\ \hline
$B^-\to \overline K^0  \overline{Z}_{uu\bar d} $&$ B^-\to K^-  \overline{Y}_{\pi u}$ &$2$                         &$B^-\to \pi^-  \overline{X'}_{ud\bar s} $&$ \overline B^0\to \pi^0  \overline{X'}_{ud\bar s}$ &$2$\\ \hline
$B^-\to \overline K^0  \overline{Z}_{uu\bar d} $&$ \overline B^0\to \overline K^0  \overline{Y}_{\pi u}$ &$4$     &$\overline B^0_s\to \pi^-  \overline{Y'}_{(u\bar u, s\bar s)d} $&$ \overline B^0_s\to \pi^0  \overline{Y'}_{(d\bar d, s\bar s)u}$ &$2$\\ \hline
$B^-\to \overline K^0  \overline{Z}_{uu\bar d} $&$ \overline B^0\to K^-  \overline{Y}_{\pi d}r$ &$4$          &$\overline B^0_s\to \overline K^0  \overline{X'}_{su\bar d} $&$ \overline B^0_s\to K^-  \overline{Y'}_{(u\bar u, d\bar d)s}$ &$2$\\ \hline
$B^-\to \eta  \overline{Z}_{uu\bar s} $&$ \overline B^0\to \eta  \overline{X}_{ud\bar s}$ &$2$             &&&  \\ \hline
$\overline B^0_s\to \pi^+  \overline{Z}_{uu\bar d} $&$ \overline B^0_s\to \pi^0  \overline{Y}_{\pi u}$ &$8(3-2 \sqrt{2})$&&&  \\ \hline
$\overline B^0_s\to \pi^+  \overline{Z}_{uu\bar d} $&$ \overline B^0_s\to \pi^-  \overline{Y}_{\pi d}$ &$4$&&&  \\ \hline
$\overline B^0_s\to \pi^-  \overline{Y}_{\pi d} $&$ \overline B^0_s\to \eta  \overline{Y}_{\pi u}$ &$6 \left(3+2 \sqrt{2}\right)$&&&  \\ \hline
$\overline B^0_s\to K^+ \overline{Z}_{uu\bar s} $&$ \overline B^0_s\to K^0  \overline{X}_{ud\bar s}$ &$2$&&&  \\ \hline
$\overline B^0_s\to \overline K^0 \overline{X}_{su\bar d} $&$ \overline B^0_s\to K^-  \overline{Y}_{\pi s}$ &$2$&&&  \\ \hline
\\\hline
$B^-\to \pi^+  Z_{dd\bar u} $&$ B^-\to K^+  X_{ds\bar u}$ &$2$                       &$B^-\to K^-  X'_{ud\bar s} $&$ B^-\to \pi^-  Y'_{(d\bar d, s\bar s)u}$ &$2$\\ \hline
$\overline B^0\to \pi^-  Z_{uu\bar d} $&$ \overline B^0\to K^0  X_{su\bar d}$ &$2$    &$\overline B^0\to \overline K^0  X'_{ud\bar s} $&$ \overline B^0\to \pi^+  Y'_{(u\bar u, s\bar s)d}$ &$2$\\ \hline
$\overline B^0_s\to K^+  Y_{\pi d} $&$ \overline B^0_s\to K^0  Y_{\pi u}$ &$1$&&&  \\ \hline
\\\hline
$B^-\to K^+  Z_{ss\bar u} $&$ B^-\to \pi^+  X_{ds\bar u}$ &$2$                        &$B^-\to \pi^-  X'_{su\bar d} $&$ B^-\to K^-  Y'_{(d\bar d, s\bar s)u}$ &$2$\\ \hline
$B^-\to \overline K^0  Y_{\pi d} $&$ \overline B^0\to \overline K^0  Y_{\pi u}$ &$1$  &$\overline B^0\to \pi^+  Y'_{(u\bar u, d\bar d)s} $&$ \overline B^0\to \pi^0  X'_{su\bar d}$ &$1$\\ \hline
$B^-\to K^-  Y_{\pi u} $&$ B^-\to \overline K^0  Y_{\pi d}$ &$2$                      &$\overline B^0\to \eta  X'_{su\bar d} $&$ B^-\to \eta  Y'_{(u\bar u, d\bar d)s}$ &$2$\\ \hline
$\overline B^0\to \pi^+  Y_{\pi s} $&$ \overline B^0\to \pi^0  X_{su\bar d}$ &$1$     &$\overline B^0_s\to \pi^+  Y'_{(u\bar u, s\bar s)d} $&$ \overline B^0_s\to \pi^0  Y'_{(d\bar d, s\bar s)u} $ &$2$\\\hline
$\overline B^0\to \pi^+  Y_{\eta s} $&$ B^-\to \pi^0  Y_{\eta s}$ &$2$                &$\overline B^0_s\to K^0  X'_{su\bar d} $&$ \overline B^0_s\to K^+  Y'_{(u\bar u, d\bar d)s}$ &$2$\\ \hline
$\overline B^0\to K^-  Z_{uu\bar d} $&$ B^-\to \overline K^0  Y_{\pi d}$ &$4$&&& \\ \hline
$\overline B^0\to \eta  X_{su\bar d} $&$ B^-\to \eta  Y_{\pi s}$ &$2$&&& \\ \hline
$\overline B^0_s\to \pi^+  Y_{\pi d} $&$ \overline B^0_s\to \eta  Y_{\pi u}$ &$6 \left(3+2 \sqrt{2}\right)$&&& \\ \hline
$\overline B^0_s\to \pi^-  Z_{uu\bar d} $&$ \overline B^0_s\to \pi^+  Y_{\pi d}$ &$4$&&& \\ \hline
$\overline B^0_s\to \pi^-  Z_{uu\bar d} $&$ \overline B^0_s\to \pi^0  Y_{\pi u}$ &$8(3-2\sqrt{2})$&&& \\ \hline
$\overline B^0_s\to K^0  X_{su\bar d} $&$ \overline B^0_s\to K^+  Y_{\pi s}$ &$2$&&& \\ \hline
$\overline B^0_s\to K^-  Z_{uu\bar s} $&$ \overline B^0_s\to \overline K^0  X_{ud\bar s}$ &$2$&&& \\ \hline
\end{tabular}
\end{table}

There are decays with  equal or proportional decay widths. This class of decays happen among $\Delta S=0$
or $\Delta S = 1$ separately which can be extracted from Tables \ref{Bc2XcS016} to \ref{B2XcS0115}.
There are several of them. For convenience, we summarize them in Table \ref{decay-rate}.
In Tables \ref{decay-rate} to \ref{U-Bc} the left three columns are for the ${\bf 15}$-plet or ${\bf \overline{15}}$-plet, while the right three columns are for the ${\bf 6}$-plet or ${\bf \bar{6}}$-plet.
In Table \ref{decay-rate} $R$ is defined as the ratio of  decay widths, 
\begin{eqnarray}
R = \Gamma_{channel-1} /\Gamma_{channel-2}
\end{eqnarray} of decay modes for ``channel 1'' and ``channel 2'' on the same line. For the pairs of such decays, all have equal CP asymmetry $A_{CP}(channel-i)$. For example, for the first pairs on the left in the table, one has
\begin{eqnarray}
\Gamma(B_c^-\to \pi^+  Z_{dd\bar u})= 2\Gamma(B_c^-\to K^+  X_{ds\bar u})\;,\;\;\;\;A_{CP}(B_c^-\to \pi^+  Z_{dd\bar u})= A_{CP}(B_c^-\to K^+  X_{ds\bar u})\;.
\end{eqnarray}

Similarly, one can read from the table for other decay modes. The relations in Table \ref{decay-rate} hold if isospin symmetry holds. They are broken by small isospin violating effects. Measurements of the above equalities can provide details information about charmed tetraquarks.

{  We comment that CP asymmetry} $A_{CP}$  in $B_c \to X_c P$ can be sizeable, of order 10\% similar to $B \to PP$, due to both tree and penguin contributions. While for $CP$ asymmetry $A_{CP}$ for $B \to \overline{ X_c} P$ and $B \to X_c P$ are much smaller since   there are only tree contributions.

\subsection{$U$-spin relations for $B\to \overline{ X}_c P$ and $B \to X_c P$ decays}

In $B\to \overline{X}_c P$ and $B \to X_c P$ decays, several decay amplitudes are related by $U$-spin symmetry, the  exchange of d and s quarks. This  leads to the ratios of decay widths proportional to CKM matrix elements ratio
$|V_{is}|^2/|V_{id}|^2$ with $i = u, c$.

For  $B \to \overline{X}_c P$ the decay amplitudes are equal to the irreducible  amplitudes multiply the CKM factor
$V_{cb}V^*_{ud}$ and $V_{cb}V_{us}^*$ for $\Delta S =0$ and $\Delta S = 1$ decay modes, respectively. Therefore
the pairs listed in Table \ref{U-spin-barXc}, the ratio of the decay widths will be given by
\begin{eqnarray}
{\Gamma(channel-1) \over \Gamma(channel-2)} = r^2 {|V_{ud}|^2\over |V_{us}|^2}\;,
\end{eqnarray}
where the parameter $r$ is defined by $r = A_{B\overline{X}_c}(channel-1)/A_{B\overline{X}_c} (channel-2)$.

\begin{table}
\caption{$U$-spin relations for $B \to \overline{X}_c P$ decays involving  both the ${\bf 6}$ and ${\bf \overline{15}}$. Results in the ``channel 1" are for $b\to d$ processes and the ones   in the ``channel 2" are for $b\to s$ processes. $r$ denotes the ratio of the two amplitudes.}\label{U-spin-barXc}
\begin{tabular}{|c|c|c||c|c|c|}\hline\hline
channel 1 & channel 2 & $r$ &channel 1 & channel 2 & $r$ \\\hline
$B^-\to K^0  \overline{Z}_{uu\bar s} $ & $ B^-\to \overline K^0  \overline{Z}_{uu\bar d}$ & $1$                                                   &$B^-\to \pi^-  \overline{Y'}_{(d\bar d, s\bar s)u} $ & $ B^-\to \pi^-  \overline{X'}_{ud\bar s}$ & $\frac{1}{\sqrt{2}}$\\ \hline
$B^-\to K^0  \overline{Z}_{uu\bar s} $ & $ B^-\to K^-  \overline{Y}_{\pi u}$ & $\sqrt{2}$                                                         &$B^-\to \pi^-  \overline{Y'}_{(d\bar d, s\bar s)u} $ & $ B^-\to K^-  \overline{Y'}_{(d\bar d, s\bar s)u}$ & $-1$\\ \hline
$B^-\to K^0  \overline{Z}_{uu\bar s} $ & $ \overline B^0\to \overline K^0  \overline{Y}_{\pi u}$ & $-2$                                           &$B^-\to \pi^-  \overline{Y'}_{(d\bar d, s\bar s)u} $ & $ \overline B^0\to \pi^0  \overline{X'}_{ud\bar s}$ & $-1$\\ \hline
$B^-\to K^0  \overline{Z}_{uu\bar s} $ & $ \overline B^0\to K^-  \overline{Y}_{\pi d}$ & $2$                                                     & $B^-\to K^-  \overline{X'}_{su\bar d} $ & $ B^-\to \pi^-  \overline{X'}_{ud\bar s}$ & $-1$\\ \hline
$B^-\to K^-  \overline{X}_{su\bar d} $ & $ B^-\to \pi^0  \overline{Z}_{uu\bar s}$ & $1$                                                          & $B^-\to K^-  \overline{X'}_{su\bar d} $ & $ B^-\to K^-  \overline{Y'}_{(d\bar d, s\bar s)u}$ & $\sqrt{2}$\\ \hline
$B^-\to K^-  \overline{X}_{su\bar d} $ & $ B^-\to \pi^-  \overline{X}_{ud\bar s}$ & $1$                                                           &$B^-\to K^-  \overline{X'}_{su\bar d} $ & $ \overline B^0\to \pi^0  \overline{X'}_{ud\bar s}$ & $\sqrt{2}$\\ \hline
$\overline B^0\to \pi^+  \overline{Z}_{uu\bar d} $ & $ \overline B^0_s\to K^+  \overline{Z}_{uu\bar s}$ & $1$                                     &$\overline B^0\to \pi^-\overline{Y'}_{(u\bar u, s\bar s)d} $ & $ \overline B^0_s\to \overline K^0 \overline{X'}_{su\bar d}$ &$-\frac{1}{\sqrt{2}}$\\\hline
$\overline B^0\to \pi^+  \overline{Z}_{uu\bar d} $ & $ \overline B^0_s\to K^0  \overline{X}_{ud\bar s}$ & $\sqrt{2}$                             & $\overline B^0\to \pi^-  \overline{Y'}_{(u\bar u, s\bar s)d} $ & $ \overline B^0_s\to K^-  \overline{Y'}_{(u\bar u, d\bar d)s}$ & $-1$\\ \hline
$\overline B^0\to K^+  \overline{Z}_{uu\bar s} $ & $ \overline B^0_s\to \pi^+  \overline{Z}_{uu\bar d}$ & $1$                                     &$\overline B^0\to K^0  \overline{X'}_{ud\bar s} $ & $ \overline B^0_s\to \overline K^0  \overline{X'}_{su\bar d}$ & $-1$\\ \hline
$\overline B^0\to K^+ \overline{Z}_{uu\bar s} $ & $ \overline B^0_s\to \pi^0  \overline{Y}_{\pi u}$ & $4-2 \sqrt{2}$                             &$\overline B^0\to K^0  \overline{X'}_{ud\bar s}$ & $ \overline B^0_s\to K^-  \overline{Y'}_{(u\bar u, d\bar d)s}$ & $-\sqrt{2}$\\ \hline
$\overline B^0\to K^+  \overline{Z}_{uu\bar s} $ & $ \overline B^0_s\to \pi^-  \overline{Y}_{\pi d}$ & $2$                                        &$\overline B^0\to \overline K^0  \overline{X'}_{su\bar d} $ & $ \overline B^0_s\to K^0  \overline{X'}_{ud\bar s}$ & $-1$\\ \hline
$\overline B^0\to K^+  \overline{Z}_{uu\bar s} $ & $ \overline B^0_s\to \eta  \overline{Y}_{\pi u}$ & $-\frac{4 \sqrt{3}}{\sqrt{2}-2}$            &$\overline B^0\to K^- \overline{Y'}_{(u\bar u, d\bar d)s} $ & $ \overline B^0_s\to \pi^0  \overline{Y'}_{(d\bar d, s\bar s)u}$ & $\sqrt{2}$\\ \hline
$\overline B^0\to K^0  \overline{X}_{ud\bar s} $ & $ \overline B^0_s\to \overline K^0  \overline{X}_{su\bar d}$ & $1$                             &$\overline B^0\to K^-  \overline{Y'}_{(u\bar u, d\bar d)s} $ & $ \overline B^0_s\to \pi^-  \overline{Y'}_{(u\bar u, s\bar s)d}$ & $-1$\\ \hline
$\overline B^0\to K^0  \overline{X}_{ud\bar s} $ & $ \overline B^0_s\to K^-  \overline{Y}_{\pi s}$ & $\sqrt{2}$                                   &$\overline B^0_s\to \pi^0  \overline{X'}_{su\bar d} $ & $ \overline B^0\to K^-  \overline{Y'}_{(u\bar u, s\bar s)d}$ & $1$\\ \hline
$\overline B^0\to \overline K^0  \overline{X}_{su\bar d} $ & $ \overline B^0_s\to K^+  \overline{Z}_{uu\bar s}$ & $\frac{1}{\sqrt{2}}$            &$\overline B^0_s\to \pi^-  \overline{Y'}_{(u\bar u, d\bar d)s} $ & $ \overline B^0\to K^-  \overline{Y'}_{(u\bar u, s\bar s)d}$ & $-1$\\ \hline
$\overline B^0\to \overline K^0  \overline{X}_{su\bar d} $ & $ \overline B^0_s\to K^0  \overline{X}_{ud\bar s}$ & $1$                             &$\overline B^0_s\to K^0  \overline{Y'}_{(d\bar d, s\bar s)u} $ & $ \overline B^0\to \overline K^0  \overline{Y'}_{(d\bar d, s\bar s)u}$ & $-1$\\  \hline
$\overline B^0_s\to K^+  \overline{Z}_{uu\bar d} $ & $ \overline B^0\to \pi^+  \overline{Z}_{uu\bar s}$ & $1$         &&&\\\hline
$\overline B^0_s\to K^0  \overline{Y}_{\pi u} $ & $ \overline B^0\to \pi^+  \overline{Z}_{uu\bar s}$ & $-\frac{1}{2}$&&&\\ \hline
$\overline B^0_s\to K^-  \overline{Z}_{ss\bar d} $ & $ \overline B^0\to \pi^-  \overline{Z}_{dd\bar s}$ & $1$&&&\\ \hline
\end{tabular}
\end{table}

\begin{table}
\caption{$U$-spin relations for $B$ decays involving  both the ${\bf \bar{6}}$ and ${\bf 15}$. Results in the ``channel 1" are for $b\to d$ processes and the ones   in the ``channel 2" are for $b\to s$ processes. $r$ denotes the ratio of the two amplitudes.}
\label{U-spin-Xc}
\begin{tabular}{|c|c|c||c|c|c|}\hline\hline
channel 1 & channel 2 & $r$ &channel 1 & channel 2 & $r$ \\\hline
$B^-\to \pi^+  Z_{dd\bar u} $ & $ B^-\to \pi^+  X_{ds\bar u}$ & $\sqrt{2}$                                            & $B^-\to \pi^-  Y'_{(d\bar d, s\bar s)u} $ & $ B^-\to \pi^-  X'_{su\bar d}$ & $-\frac{1}{\sqrt{2}}$\\ \hline
$B^-\to \pi^+  Z_{dd\bar u} $ & $ B^-\to K^+  Z_{ss\bar u}$ & $1$                                                     & $B^-\to \pi^-  Y'_{(d\bar d, s\bar s)u} $ & $ B^-\to K^-  Y'_{(d\bar d, s\bar s)u}$ & $-1$\\ \hline
$B^-\to K^+  X_{ds\bar u} $ & $ B^-\to \pi^+  X_{ds\bar u}$ & $1$                                                      &$B^-\to K^+  X'_{ds\bar u} $ & $ B^-\to \pi^+  X'_{ds\bar u}$ & $-1$\\ \hline
$B^-\to K^+  X_{ds\bar u} $ & $ B^-\to K^+  Z_{ss\bar u}$ & $\frac{1}{\sqrt{2}}$                                      & $B^-\to K^0  Y'_{(u\bar u, d\bar d)s} $ & $ B^-\to \overline K^0  Y'_{(u\bar u, s\bar s)d}$ & $-1$\\ \hline
$B^-\to \overline K^0  Z_{dd\bar s} $ & $ B^-\to K^0  Z_{ss\bar d}$ & $1$                                             & $B^-\to K^-  X'_{ud\bar s} $ & $ B^-\to \pi^-  X'_{su\bar d}$ & $-1$\\ \hline
$B^-\to K^-  X_{ud\bar s} $ & $ B^-\to \pi^-  X_{su\bar d}$ & $1$                                                     & $B^-\to K^-  X'_{ud\bar s} $ & $ B^-\to K^-  Y'_{(d\bar d, s\bar s)u}$ & $-\sqrt{2}$\\ \hline
$\overline B^0\to \pi^-  Z_{uu\bar d} $ & $ \overline B^0_s\to \overline K^0  X_{ud\bar s}$ & $\sqrt{2}$               &$\overline B^0\to \pi^+  Y'_{(u\bar u, s\bar s)d} $ & $ \overline B^0_s\to K^+  Y'_{(u\bar u, d\bar d)s}$ & $-1$\\ \hline
$\overline B^0\to \pi^-  Z_{uu\bar d} $ & $ \overline B^0_s\to K^-  Z_{uu\bar s}$ & $1$                               & $\overline B^0\to \pi^+  Y'_{(u\bar u, s\bar s)d} $ & $ \overline B^0_s\to K^0  X'_{su\bar d}$ & $-\frac{1}{\sqrt{2}}$\\ \hline
$\overline B^0\to K^0  X_{su\bar d} $ & $ \overline B^0_s\to \overline K^0  X_{ud\bar s}$ & $1$                      & $\overline B^0\to K^+  Y'_{(u\bar u, d\bar d)s} $ & $ \overline B^0_s\to \pi^+  Y'_{(u\bar u, s\bar s)d}$ & $-1$\\ \hline
$\overline B^0\to K^0  X_{su\bar d} $ & $ \overline B^0_s\to K^-  Z_{uu\bar s}$ & $\frac{1}{\sqrt{2}}$                & $\overline B^0\to K^+  Y'_{(u\bar u, d\bar d)s} $ & $ \overline B^0_s\to \pi^0  Y'_{(d\bar d, s\bar s)u}$ & $\sqrt{2}$\\ \hline
$\overline B^0\to \overline K^0  X_{ud\bar s} $ & $ \overline B^0_s\to K^+  Y_{\pi s}$ & $\sqrt{2}$                   & $\overline B^0\to K^0  X'_{su\bar d} $ & $ \overline B^0_s\to \overline K^0  X'_{ud\bar s}$ & $-1$\\ \hline
$\overline B^0\to \overline K^0  X_{ud\bar s} $ & $ \overline B^0_s\to K^0  X_{su\bar d}$ & $1$                      &$\overline B^0\to \overline K^0  X'_{ud\bar s} $ & $ \overline B^0_s\to K^+  Y'_{(u\bar u, d\bar d)s}$ & $-\sqrt{2}$\\ \hline
$\overline B^0\to K^-  Z_{uu\bar s} $ & $ \overline B^0_s\to \pi^+  Y_{\pi d}$ & $2$                                  & $\overline B^0\to \overline K^0  X'_{ud\bar s} $ & $ \overline B^0_s\to K^0  X'_{su\bar d}$ & $-1$\\ \hline
$\overline B^0\to K^-  Z_{uu\bar s} $ & $ \overline B^0_s\to \pi^0  Y_{\pi u}$ & $4-2 \sqrt{2}$                        &$\overline B^0_s\to K^+  Y'_{(u\bar u, s\bar s)d} $ & $ \overline B^0\to \pi^+  Y'_{(u\bar u, d\bar d)s}$ & $-1$\\ \hline
$\overline B^0\to K^-  Z_{uu\bar s} $ & $ \overline B^0_s\to \pi^-  Z_{uu\bar d}$ & $1$                                &$\overline B^0_s\to K^+  Y'_{(u\bar u, s\bar s)d} $ & $ \overline B^0\to \pi^0  X'_{su\bar d}$ & $1$\\ \hline
$\overline B^0\to K^-  Z_{uu\bar s} $ & $ \overline B^0_s\to \eta  Y_{\pi u}$ & $-\frac{4 \sqrt{3}}{\sqrt{2}-2}$      & $\overline B^0_s\to K^0  Y'_{(d\bar d, s\bar s)u} $ & $ \overline B^0\to \overline K^0  Y'_{(d\bar d, s\bar s)u}$ & $-1$\\ \hline
$\overline B^0_s\to \pi^+  Z_{dd\bar s} $ & $ \overline B^0\to K^+  Z_{ss\bar d}$ & $1$&&&\\ \hline
$\overline B^0_s\to \pi^-  Z_{uu\bar s} $ & $ B^-\to \overline K^0  Y_{\pi d}$ & $2$&&&\\ \hline
$\overline B^0_s\to \pi^-  Z_{uu\bar s} $ & $ B^-\to K^-  Y_{\pi u}$ & $\sqrt{2}$&&&\\ \hline
$\overline B^0_s\to \pi^-  Z_{uu\bar s} $ & $ \overline B^0\to \overline K^0  Y_{\pi u}$ & $-2$&&&\\ \hline
$\overline B^0_s\to \pi^-  Z_{uu\bar s} $ & $ \overline B^0\to K^-  Z_{uu\bar d}$ & $1$&&&\\ \hline
\end{tabular}
\end{table}

For  $B \to  X_c P$ the decay amplitudes are equal to the irreducible  amplitudes multiply the CKM factor
$V_{ub}V^*_{cd}$ and $V_{ub}V_{cs}^*$ for $\Delta S =0$ and $\Delta S = 1$ decay modes, respectively. Thereby for
the pairs listed in Table \ref{U-spin-barXc}, the ratio of the decay widths will be given by
\begin{eqnarray}
{\Gamma(channel-1) \over \Gamma(channel-2)} = r^2 {|V_{cd}|^2\over |V_{cs}|^2}\;.
\end{eqnarray}
In this case the parameter $r$ is defined by $r = A_{BX_c}(channel-1)/A_{BX_c} (channel-2)$. 
We give  the  coefficient  $r$  for $B\to \overline{X}_c P$ and $B\to X_c P$ in
Tables \ref{U-spin-barXc} and \ref{U-spin-Xc}. One can read off the relations for pairs related by $U$-spin and test  the  symmetry  by  comparing the relations  with data in future.

\subsection{$U$-spin for $B_c \to X_c P$ decay and CP violating relations}

\begin{table}
\caption{$U$-spin relations for $B_c$ decays involving  both the ${\bf \bar{6}}$-plet and ${\bf 15}$-plet. Results in the ``channel 1" are for $b\to d$ processes and the ones   in the ``channel 2" are for $b\to s$ processes. $r$ denotes the ratio of the two amplitudes.}
\label{U-Bc}
\begin{tabular}{|c|c|c||c|c|c|}\hline\hline
channel 1 & channel 2 & $r$ &channel 1 & channel 2 & $r$ \\\hline
  $B_c^-\to \pi^+  Z_{dd\bar u}$ & $ B_c^-\to \pi^+  X_{ds\bar u}$ & $\sqrt{2} $& $B_c^-\to \pi^-  Y'_{(d\bar d, s\bar s)u}$ & $ B_c^-\to \pi^-  X'_{su\bar d}$ & $-\frac{1}{\sqrt{2}} $   \\ \hline
 $B_c^-\to \pi^+  Z_{dd\bar u}$ & $ B_c^-\to K^+  Z_{ss\bar u}$ & $1 $&$B_c^-\to \pi^-  Y'_{(d\bar d, s\bar s)u}$ & $ B_c^-\to K^-  Y'_{(d\bar d, s\bar s)u}$ & $-1 $         \\ \hline
  $B_c^-\to K^+  X_{ds\bar u}$ & $ B_c^-\to \pi^+  X_{ds\bar u}$ & $1 $&$B_c^-\to K^+  X'_{ds\bar u}$ & $ B_c^-\to \pi^+  X'_{ds\bar u}$ & $-1 $     \\ \hline
   $B_c^-\to K^+  X_{ds\bar u}$ & $ B_c^-\to K^+  Z_{ss\bar u}$ & $\frac{1}{\sqrt{2}} $&$B_c^-\to K^0  Y'_{(u\bar u, d\bar d)s}$ & $ B_c^-\to \overline K^0  Y'_{(u\bar u, s\bar s)d}$ & $-1 $\\ \hline
$B_c^-\to \overline K^0  Z_{dd\bar s}$ & $ B_c^-\to K^0  Z_{ss\bar d}$ & $1 $&$B_c^-\to K^-  X'_{ud\bar s}$ & $ B_c^-\to \pi^-  X'_{su\bar d}$ & $-1 $\\ \hline
  $B_c^-\to K^-  X_{ud\bar s}$ & $ B_c^-\to \pi^-  X_{su\bar d}$ & $1 $&$B_c^-\to K^-  X'_{ud\bar s}$ & $ B_c^-\to K^-  Y'_{(d\bar d, s\bar s)u}$ & $-\sqrt{2} $\\ \hline
\end{tabular}
\end{table}

For $B_c \to X_c P$ decays, there are two terms with different CKM factors. Although the $U$-spin symmetry can relate 
$\Delta S=0$ and $\Delta S =1$ decays, there is no simple rate relation  as in the case  for $B \to \overline{X_c} P$ and $B \to X_c P$ decays.
However, there exists a relation for the CP violating quantity
$\Delta = \Gamma - \bar \Gamma$. 
We now derive such a relation. Let us consider two $U$-spin connected   decays with proportional amplitudes $A^T_{B_c}$ and $A^P_{B_c}$, with the decay amplitudes 
\begin{eqnarray}
&&A(\Delta S = 0) = r\left (V_{ub}V_{ud}^*A^T_{B_c} + V_{tb}V^*_{td} A^P_{B_c}\right )\;,\nonumber\\
&&A(\Delta S = 1) = V_{ub}V_{us}^*A^T_{B_c} + V_{tb}V^*_{ts} A^P_{B_c}\;.
\end{eqnarray}
Using the Jarlskog relation
$Im(V_{ub}V_{ud}^*V^*_{tb}V_{td}) = -  Im(V_{ub}V_{us}^*V^*_{tb}V_{ts})$,
the CP violating rate difference $\Delta (\Delta S = i) = \Gamma(\Delta S= i) - \overline{\Gamma}(\Delta S= i)$ will be related with~\cite{Deshpande:1994ii,He:1998rq,Gronau:2000zy}
\begin{eqnarray}
r^2 \Delta(\Delta S=0) = -\Delta (\Delta S = 1)\;.
\end{eqnarray}
This leads to branching ratio ${\cal B}(\Delta S = i)$ and CP asymmetry $A_{CP}(\Delta S= i)$ relation,
\begin{eqnarray}
{A_{CP}(\Delta S=0)\over A_{CP}(\Delta S = 1)}= - r^2{{\cal B}(\Delta S = 1)\over {\cal B}(\Delta S =0)}\;.\label{CPR}
\end{eqnarray}

In Table \ref{U-Bc}, we collect the $B_c$ decay pairs related by $U$-spin. CP asymmetries for these pairs satisfy relation in Eq.(\ref{CPR}).
As already mentioned,  CP asymmetries for $B_c\to X_c P$ decays are expected to be similar to $B\to PP$ decays which can be of order 10\%. Experimental measurements of these  relations are important to test flavor $SU(3)$ symmetry and also CKM mechanism for CP violation.

 \section{conclusions}
 \label{sec:conclusion}

In this work we have studied the charmed tetraquarks with three light flavors  in  weak decays of $B_c$ and $B$ mesons. If indeed a charmed tetraquark $X_c$ with three different light quarks is discovered, they should come in a ${\bf \bar 6}$ or a ${\bf 15}$ multiplet of flavor $SU(3)$. A most direct consequence  of the flavor $SU(3)$ symmetry is  that tetraquarks are formed in irreducible representations. Therefore 
to test whether flavor $SU(3)$ symmetry is playing a role in organizing hadron states, one needs to find all of the states in a given multiplet.  We have studied production of $X_c$ in $B_c$ and $B$ weak decays. The total number of states with similar masses will be able to distinguish whether $X_c$ is in ${\bf \bar 6}$ or ${\bf 15}$. The doubly charged states $Z_{dd\bar u}$ and $Z_{ss\bar u}$ are the smoking guns for $X_c$ belonging  to ${\bf 15}$. We   find a number of  relations among decay branching ratio for $\Delta S =0$ and $\Delta S=1$ processes separately. We also find relations among the CP asymmetries and branching ratios for decays related by the $U$-spin symmetry. All these can serve to confirm the existence of tetraquark states and to study their properties. With more data from the LHCb and Belle-II, one may discover  the $X_c$ states and study their properties. We urge our experimental colleagues to carry out related analysis to learn  more about hadron states built from multiquarks.

\section*{Acknowledgement}

This work was supported in part by  National Natural Science
Foundation of China  under Grant  No.11575110, 11575111, by Natural  Science Foundation of Shanghai under Grant  No.15DZ2272100, No. 15ZR1423100, by China Postdoctoral Science Foundation,  by the Open Project Program of State Key Laboratory of Theoretical Physics, Institute of Theoretical Physics, Chinese  Academy of Sciences, China (No.Y5KF111CJ1), and  by   Scientific Research Foundation for  Returned Overseas Chinese Scholars, State Education Ministry. X.~G. He was also supported in part by MOE Academic Excellent Program (Grant No.~102R891505), NCTS and MOST of ROC (Grant No.~MOST104-2112-M-002-015-MY3). X.~G. He thanks Korea Institute for Advanced Study (KIAS) for their hospitality and partial support when this work was completed.

\end{document}